\documentclass[11pt,a4paper]{article}

\usepackage[margin=2.5cm]{geometry}
\usepackage[utf8]{inputenc}
\usepackage{fontenc,authblk,graphicx}

\usepackage{amsmath,amssymb,mathrsfs,amsfonts,bm}
\usepackage[dvipsnames]{xcolor}
\usepackage[colorlinks=true,citecolor=blue,urlcolor=blue,linkcolor=blue]{hyperref}

\usepackage{subcaption,braket}

\usepackage[style=phys,hyperref=true,articletitle=true,biblabel=brackets,chaptertitle=false]{biblatex}
\def \be {\begin{equation}} 
\def \ee {\end{equation}} 
\def \l {\left(} 
\def \r {\right)} 
\def \la {\langle} 
\def \ra {\rangle}

\title{\bf A hydrodynamic approach to Stark localization}

\author[1]{Luca Capizzi\thanks{lcapizzi@sissa.it}}
\author[1]{Carlo Vanoni\thanks{cvanoni@sissa.it}}
\author[1,2]{Pasquale Calabrese}
\author[1]{Andrea Gambassi}

\affil[1]{SISSA -- International School for Advanced Studies and INFN, via Bonomea 265, 34136, Trieste, Italy}
\affil[2]{The Abdus Salam ICTP, Strada Costiera 11, 34151, Trieste, Italy}

\date{\today}

\begin{document}

\maketitle
\begin{abstract}

When a free Fermi gas on a lattice is subject to the action of a linear potential it does not drift away, as one would naively expect, but it remains spatially localized.
Here we revisit this phenomenon, known as Stark localization, within the recently proposed framework of generalized hydrodynamics.
In particular, we consider the dynamics of an initial state in the form of a domain wall and we recover known results for the particle density and the particle current, while we derive analytical predictions for relevant observables such as the entanglement entropy and the full counting statistics.
Then, we extend the analysis to generic potentials, highlighting the relationship between the occurrence of localization and the presence of peculiar closed orbits in phase space, arising from the lattice dispersion relation. We also compare our analytical predictions with numerical calculations and with the available results, finding perfect agreement. This approach paves the way for an exact treatment of the interacting case known as Stark many-body localization.
\end{abstract}

\section{Introduction}

Understanding and characterizing the dynamics of quantum many-body systems is one of the central themes of modern physics. For instance, given an initial state of an isolated system, which is left to evolve unitarily, one is interested in the time evolution of local observables. Generically one expects local relaxation to thermal ensembles~\cite{pssv-11} to occur. However, it has been shown that the stationary states of integrable systems are actually described by a generalized Gibbs ensemble (GGE)~\cite{rdyo-07,sc-14,vr-16,ef-16}, due to the presence of an extensive number of conservation laws. A systematic theoretical approach to investigate the dynamics of inhomogeneous integrable systems, including in particular free theories, has been recently formulated in the form of a generalized hydrodynamics (GHD)~\cite{cadt-16,bcdnf-16}. This approach extends standard hydrodynamics by accounting for the additional conservation laws enforced by integrability.
GHD turned out to be a versatile and predictive method in a large variety of contexts, including transport phenomena in spin-chains~\cite{f-17,f-20,bfpc-18,sh-22,wck-13,bp-18,scd-22,pdncbf-17,cdlv-18,abfpr-21,bvkm-17,bpp-20,rc-23}, inhomogenous quantum gases both in and out of equilibrium~\cite{asw-22,bpc-18,dt-17,rcdd-21,skcd-21,fbgvp-21,kmm-18,bd-22,bdd-20,dndmp-22}, quantum and diffusion effects~\cite{agv-19,mdny-20,dnbd-19,dnbd-18,ddldnd-21,bgi-21,bdndl-20,bdlv-21,rcdd-20}, as reviewed in Refs.~\cite{e-22,d-20,sk-23}.
Its theoretical predictions have also been confirmed in recent experiments~\cite{mzldrw-21,sbdd-19}.

An early counter-intuitive discovery concerning the dynamics of non-interacting quantum particles on a lattice (described by the tight-binding model)~\cite{gp-13} and subject to a constant force was the presence of Bloch oscillations~\cite{b-29}. 
Indeed, contrary to what one may heuristically expect, it was shown that these particles display a periodic motion~\cite{w-62,hkkm-04,gp-13} instead of drifting forever.
The occurrence of this phenomenon, nowadays known as Stark localization,
is not limited to tight-binding non-interacting models, but occurs also in interacting systems. For example, this has been recently demonstrated experimentally in a 5-qubit superconducting processor \cite{gghzy-21}. In addition, it has been argued that Stark localization is robust against the presence of interaction, leading to the notion of Stark many-body localization~\cite{shmp-19,nbr-19,dgp-21} which has been observed in an experiment with a trapped-ion quantum simulator~\cite{mlbcfkpygm-21}.
Similarly, the effective dynamics of quantum collective excitations may feature Stark localization, leading to confinement~\cite{lsmpcg-20,mplcg-19}.
In spite of the evidences mentioned above, a general theoretical framework for understanding Stark localization beyond the cases of simple analytically solvable models and approximated descriptions~\cite{wck-13,lzsg-19} seems still to be missing.
In this work, we aim at partially filling this gap within the GHD approach. In particular, we shed light on a crucial question concerning Stark localization, i.e., its fate when the external potential is not linear.
We show that the occurrence of localization is associated with the existence of closed trajectories in phase space, which encircle the Brillouin zone. This does not actually require a fine-tuning of the potential but it crucially depends on the form of the lattice dispersion relation, which differs from the one on the continuum.

The rest of the presentation is organized as follows. 
In Sec.~\ref{sec:Generic} we briefly review the GHD approach, with particular emphasis on lattice Fermi gases. In Sec.~\ref{sec:GHD_Stark} we focus on the dynamics of a domain-wall initial state in the presence of a linear potential, providing analytical predictions for the particle density and current. In addition, by employing the recently proposed \emph{quantum} GHD~\cite{rcdd-20}, we investigate the evolution of the entanglement entropy and the full-counting statistics. In Sec.~\ref{sec:Stark_generic} we consider the case of generic external potentials, in order to understand which ingredients are important for the occurrence of Stark localization. We summarize our findings in Sec.~\ref{sec:Conclusion}, listing some open questions.

\section{Generalized hydrodynamics of inhomogeneous systems}
\label{sec:Generic}

In this section, we briefly review the generalized hydrodynamics, setting the stage for our investigation of the problem of Stark localization.
We consider a lattice Fermi gas with nearest-neighbor hopping, subject to an external potential $V(x)$. The corresponding Hamiltonian is
\be
\label{eq:hamiltonian}
    H = - \frac{1}{2} \sum_{x\in {\mathbb Z}} (\psi_{x}^{\dagger} \psi_{x+1} + \psi_{x+1}^{\dagger} \psi_{x}) + \sum_{x\in {\mathbb Z}} V(x) \, \psi_{x}^{\dagger} \psi_{x},
\ee
where $\psi_x$ and $\psi^\dagger_x$ are the annihilation/creation fermionic operators satisfying the canonical 
anti-commutation relations
\be
\{ \psi_x,\psi^\dagger_{x'}\} = \delta_{xx'}\quad \mbox{and}\quad \{ \psi_x,\psi_{x'}\} = 0.
\ee
Given an initial state $\ket{\Psi_0}$ and an observable $\mathcal{O}$, one is usually interested in investigating the time evolution of the expectation value of $\mathcal{O}$, i.e., of
\be
\la \mathcal{O}(t)\ra \equiv \bra{\Psi_0}e^{iHt}\mathcal{O}e^{-iHt}\ket{\Psi_0}.
\ee
Remarkably, for the large class of Gaussian initial states, one can reconstruct the evolution of any observable $\mathcal{O}$ on the basis of the two-point function only, namely
\be\label{eq:Cmatrix}
C(x,x';t) \equiv \la \psi_{x}^{\dagger}(t)\psi_{x'}(t)\ra;
\ee
this allows a drastic simplification of the treatment of the microscopic dynamics.
More generally, predicting the time evolution of the system requires the exact determination of the single-particle spectrum, which might be hard to calculate explicitly. However, it has been demonstrated that a somehow simpler hydrodynamic regime (known as inhomogeneous GHD~\cite{skcd-21}) emerges at large scales.
For instance, if the potential $V(x)$ is a sufficiently smooth function of $x$ and the multi-point correlations in the initial state decay rapidly upon increasing their distances~\cite{bdl-18}, a viable semi-classical description of the dynamics can be done in terms of a local Fermi occupation function $n(x,k;t)$ defined as~\cite{w-32,cg-69}
\be
n(x,k;t) \equiv\! \int dy\, \la \psi^\dagger_{x+y/2}(t) \psi_{x-y/2}(t) \ra e^{iky}.
\ee
This description amounts at studying the Liouville evolution to lowest order in $\partial_x$ and $\partial_k$, given by \cite{dt-17,abfpr-21,f-17}
\be
\label{eq:evolut_n}
    \partial_t n(x,k;t) + v(k)\, \partial_x n(x,k;t) =  V'(x) \, \partial_k n(x,k;t) \quad\mbox{where}\quad  v(k) = \sin k.
\ee
Here, $n(x,k;t)$ can be interpreted as a semi-classical probability distribution in the phase-space $(x,k) \in \mathbb{R} \times \mathbb{R} \mbox{\,mod\,} 2\pi$ associated to the classical Hamiltonian
\be
\label{eq:Ham_V}
    \mathcal{H}(x,k) = - \cos{k} + V(x),
\ee
which results in the following equations of motion: 
\begin{align}
\begin{cases}
    \dot{x} = {\displaystyle \frac{\partial \mathcal{H}}{\partial k}}= \sin{k},\\[3mm]
    \dot{k} = {\displaystyle -\frac{\partial \mathcal{H}}{\partial x}} = \, -V'(x).
\end{cases}
\label{eq:eq-motion}
\end{align}
Let us mention that Eq.~\eqref{eq:evolut_n} comes from the requirement that the semi-classical probability $n(x,k;t)$ is conserved (see also Eq.~\eqref{eq:conservation_number}).
Within this approach, some local observables can be directly expressed and computed in terms of $n(x,k;t)$ alone. 
In particular, the density of fermions takes the form
\be\label{eq:density}
\rho(x,t) \equiv   \la \psi^\dagger_x(t)\psi_x(t)\ra= \int^{\pi}_{-\pi}\!\frac{dk}{2\pi}\, n(x,k;t),
\ee
and the particle current
\be
    \label{eq:current}
    j(x,t) \equiv \frac{i}{2} \la  \psi_{x+1}^{\dagger}(t) \psi_{x}(t) - \psi_{x}^{\dagger}(t) \psi_{x+1}(t)\ra = \int^{\pi}_{-\pi} \!\frac{dk}{2\pi}\, n(x,k;t)v(k).
\ee
Note the following crucial point: while the hydrodynamic approach is expected to be predictive at spatial and temporal scales much larger than the microscopic ones, the presence of a lattice makes the momentum $k$ a compact variable, which is defined up to $k\rightarrow k+2\pi$. Accordingly, the lattice strongly affects the resulting dispersion relation, i.e., the form of $v(k)$ in Eq.~\eqref{eq:evolut_n} or, equivalently, the kinetic term in Eq.~\eqref{eq:Ham_V}.
In fact, after reinstating the lattice spacing $a$ in the definition of $v(k)$, one readily realizes that for $k$ smaller than $a^{-1}$ it is legitimate to  approximate
\be
v(k) = a^{-1}\sin (a k) \simeq k,
\ee
retrieving the usual Galilean dispersion. However, this is no longer the case for generic values of $k$ and the fact that $v(k)$ is a periodic function of $k$ plays a crucial role.
As anticipated, this effect of the lattice is precisely the origin of Stark localization, as we shall demonstrate in the following sections.

\section{Dynamics in the presence of a linear potential}
\label{sec:GHD_Stark}

In this section, we analyze in detail the dynamics of the standard setup in which Stark localization occurs~\cite{gp-13}, i.e., a tight-binding model of a lattice Fermi gas in the presence of a linear potential
\be
V(x) = -hx,
\label{eq:lin-pot}
\ee
where, without loss of generality, we assume $h>0$. 
The single-particle spectrum of the microscopic model has been determined exactly in Refs.~\cite{w-62,hkkm-04,gkk-02} and it features a Wannier-Stark ladder of energy levels and exponentially localized wave functions.
Moreover, as discussed in Refs.~\cite{bglsv-22,bglsv-23}, a large-scale limit of this dynamics turns out to exist for a generic initial state and in the presence of weak field $h$, i.e., with
\be
h \ll 1,
\ee
(in lattice spacing units). Correspondingly, Bloch oscillations for the density and the current starting from a domain-wall state were established analytically. In fact, as dimensional analysis suggests, $1/h$ is a length (which turns out to be a localization length, see, c.f., Eq.~\eqref{eq:loc-length}) and a semi-classical regime is expected to emerge when this length is much larger than the lattice spacing. While these previous results were derived on the basis of an explicit solution of the microscopic model, as far as we know, GHD has never been applied to this problem, which is precisely the goal of this work.

\begin{figure}[t]
    \centering
	\includegraphics[width=0.67\linewidth]{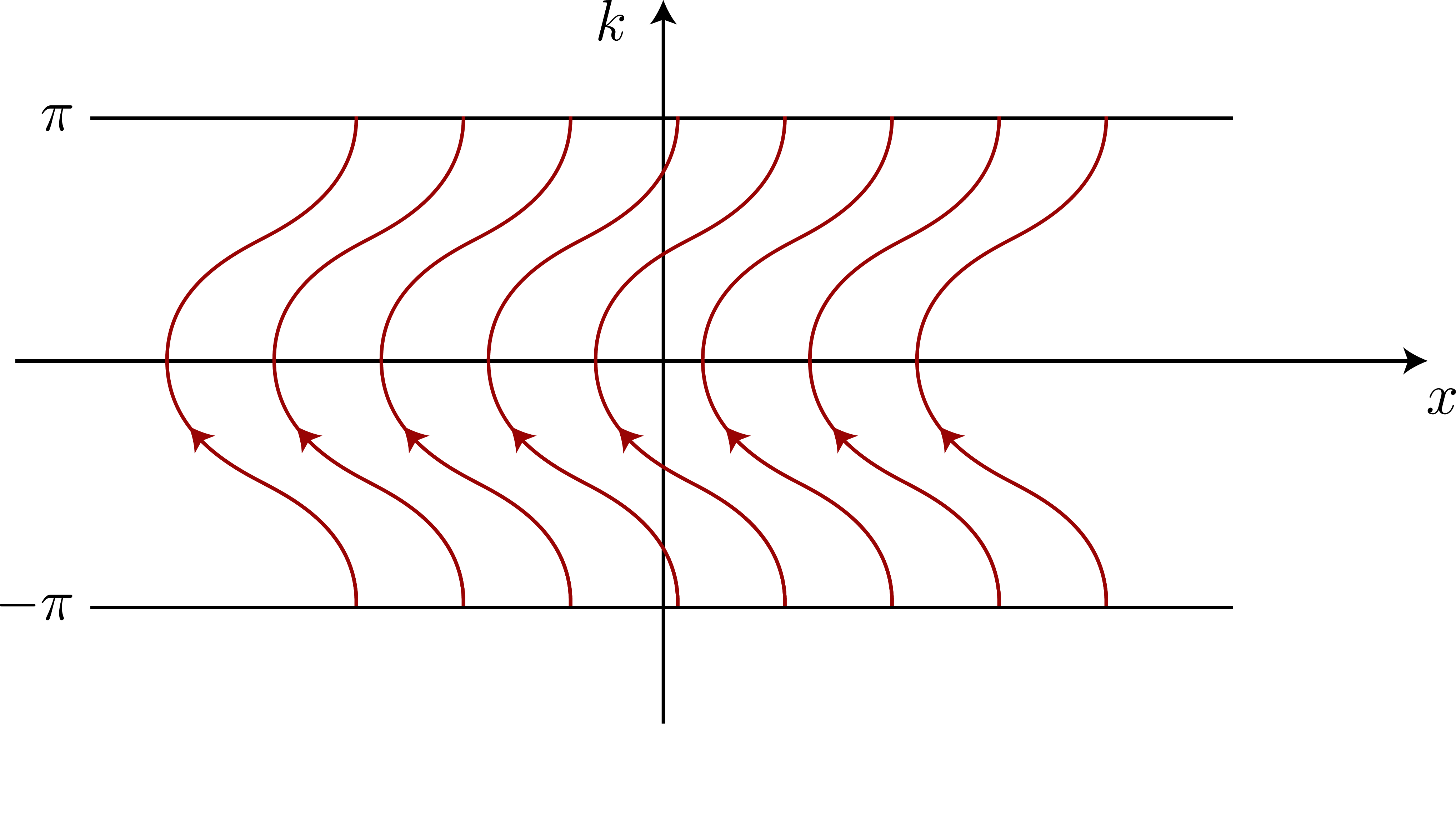}
    \caption{Classical trajectories of a particle with Hamiltonian \eqref{eq:Ham_V} in the presence of a linear potential $V(x) = -hx$, given by Eq.~\eqref{eq:xk-tevo}. Each trajectory encircles the first Brillouin zone in a period $T = 2\pi/h$.}
    \label{fig:Trajectories}
\end{figure}

Before presenting the calculation of the exact semi-classical dynamics, it is worth giving a simple physical description of the system. Let us consider a (classical) particle with the Hamiltonian \eqref{eq:Ham_V} and the potential \eqref{eq:lin-pot}, i.e., with
\be
\label{eq:Class_Ham}
\mathcal{H}(x,k) = - \cos{k} - h x,
\ee
(see also Appendix~\ref{app:sec:hamilt}) initially localized at $x(0)=0$ and $k(0)=0$. 
At short times, the particle is accelerated to the right and thus its momentum $k$ increases linearly, as one would expect in the continuum limit where the lattice is absent and the ``kinetic term'' $-\cos k \simeq k^2/2 + \mbox{const.}$ reproduces the usual one.
However, the velocity $\dot{x}$ [see Eq.~\eqref{eq:eq-motion}] does not grow indefinitely (being bounded by $|\dot{x}|\leq  1$) and the lattice provides negative feedback, slowing down the particle until it stops at position $x_1$, corresponding to the inversion point. Then, the particle is accelerated again towards the left and eventually reaches the initial position. This process is then repeated, leading to an oscillatory motion between two extreme points $x=0$ and $x=x_1$.
The value of $x_1$ is easily determined by energy conservation, requiring that the velocity at that point vanishes, finding
\be
x_1 = 2/h.
\ee
The classical trajectory in phase space $(x,k) \in \mathbb{R} \times \mathbb{R} \mbox{\,mod\,} 2\pi$ can be calculated by solving the equations of motion \eqref{eq:eq-motion} for the linear potential \eqref{eq:lin-pot}, which read
\be
\begin{cases}
    \dot{x} = \sin k,\\
    \dot{k} = h,
\end{cases}
\ee
for a generic initial condition $(x(0),k(0))$. The corresponding solution is (see, e.g., Ref.~\cite{bglsv-23})
\be
\begin{cases}
\label{eq:xk-tevo}
    k(t) = k(0) + ht, \\
    \displaystyle{x(t) = x(0) + \int_{0}^{t} dt' \, \sin{(k(0)+ht)} = x(0) + \frac{2}{h} \sin\left( k(0) + \frac{ht}{2} \right) \sin\left(\frac{ht}{2} \right).}
\end{cases}
\ee
Notice that for small $k(0)$ and $t$, i.e., $k(0)\ll 1$ and $t\ll 1/h$ one gets
\be
x(t)\simeq x(0) + k(0)t +\frac{ht^2}{2},
\ee
which, as expected, is the motion of a uniformly accelerated classical particle. Still, at longer times Eq.~\eqref{eq:xk-tevo} implies that one always observes oscillatory motion in 
$x \in [x(0)-2/h,x(0)+2/h]$, no matter how small $h\neq 0$ is. Correspondingly,  $k(t)$ periodically encircles the first Brillouin zone with a period given by
\be
\label{eq:expT}
T=\frac{2\pi}{h}.
\ee
Figure \ref{fig:Trajectories} shows the foliation of the phase space $(x,k)$ provided by the trajectories of $\mathcal{H}(x,k)$ in Eq.~\eqref{eq:xk-tevo}.

For the sake of completeness, we finally write the explicit expression of the dynamics of the probability distribution $n(x,k;t)$ in phase space for a given initial distribution $n(x,k;t=0)$ of the non-interacting particles. In order to do so, it is sufficient to rewrite the equation of motion \eqref{eq:evolut_n} as the conservation of the probability along the flow in phase space induced by the Hamiltonian \eqref{eq:Class_Ham}, i.e.,
\be
\label{eq:conservation_number}
\frac{d}{dt} n(x(t),k(t);t) =0.
\ee
In other words, by evolving backward in time the trajectory starting from $(x,k)$ one easily gets the local occupation at time $t$ from the initial one. Using Eqs.~\eqref{eq:xk-tevo} we conclude that
\be
\label{eq:solution_n}
    n(x,k;t) = n\left(x - \frac{2}{h} \sin\left(k - \frac{ht}{2}\right)\sin\left(\frac{ht}{2}\right), k- ht;0 \right),
\ee
which satisfies Eq.~\eqref{eq:evolut_n} as one can easily check.

\subsection{Domain-wall initial state}

\begin{figure}[t]
    \centering
	\includegraphics[width=0.67\linewidth]{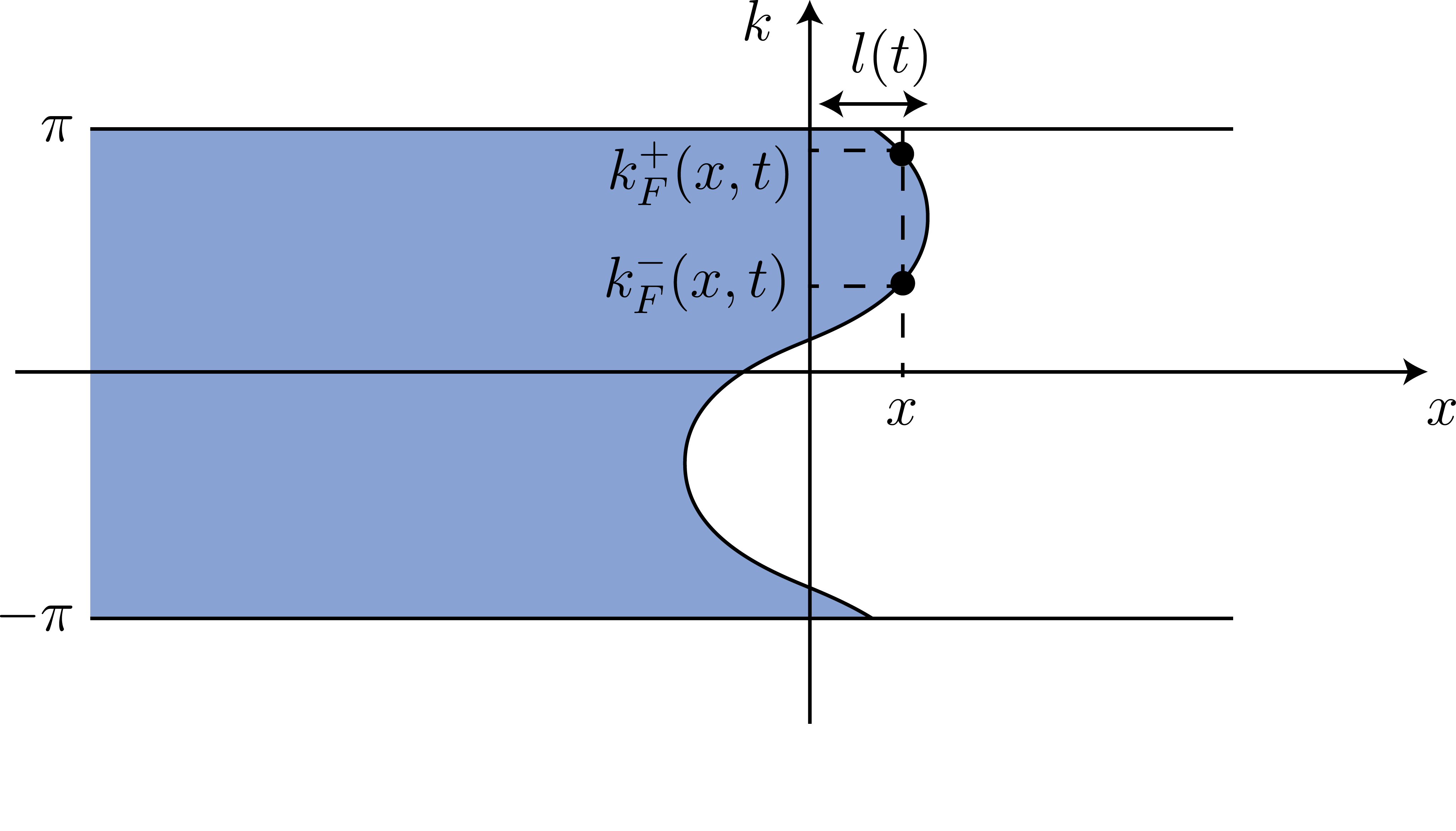}
    \caption{Local occupation $n(x,k;t)$ in phase space at given time $t>0$: $n(x,k;t) =1$ in the blue region and $n(x,k;t) =0$ in the complementary white one. For $|x|<l(t)$ there is a pair of Fermi points, denoted by $k^{\pm}_{F}(x,t)$. In the region $x>l_{\rm loc}$ ($x<-l_{\rm loc}$), with $l_{\rm loc}$ given in Eq.~\eqref{eq:l-loc}, no local evolution occurs and the system is locally described by a completely empty (filled) Fermi sea.}
    \label{fig:Phase_space}
\end{figure}

Considering now the dynamics of the quantum system, we focus 
on an initial state with a single domain wall, which has been the subject of many studies~\cite{arrs-99,akr-08,em-18,scd-22,sk-23}, and we aim at characterizing its evolution. To do so, let us first introduce the empty or vacuum state $\ket{0}$  defined by 
\be
\psi_x \ket{0} = 0, \quad \forall x\in {\mathbb Z}.
\ee
In terms of $\ket{0}$, the domain-wall state $\ket{\Psi_0}$ can be expressed as
\be\label{eq:dwall_state}
\ket{\Psi_0} = \prod_{x\leq 0} \psi_x^{\dagger}\ket{0},
\ee
and corresponds to having all lattice sites filled by one fermion for $x\leq 0$ and empty for $x>0$. 
As shown, e.g., in Ref.~\cite{scd-22}, the state $\ket{\Psi_0}$ admits a semi-classical description with local occupation given by
\be
\label{eq:domain_wall_init_state}
    n(x,k;0) = 
    \begin{cases}
    1      & \mbox{for } x \leq 0 \mbox{ and } k\in[-\pi,\pi], \\[1mm]
    0      & \mbox{otherwise}.
    \end{cases}
\ee
In particular, a Fermi contour at $x=0$ separates the phase space into an empty region and a filled one.
We now study the dynamics of $\ket{\Psi_0}$. A convenient way to express $n(x,k;t)$, which overcomes the possible issues due to its discontinuities as a function of $x$ and $k$, is via its Fermi contour (see also Ref.~\cite{scd-22}). 
For instance, the set of points $\{(x=0,k=k_0)\}$ of the initial Fermi surface, parameterized by the initial momentum $k_0 \in [-\pi,\pi)$, evolves in a time $t$ to the set
\be\label{eq:Fermi_prof}
\left\{(x_t,k_t) \ | \
x_t = \frac{2}{h}\sin\l \frac{h t }{2}\r \sin\l k_t -\frac{h t }{2}\r  \quad \mbox{with} \quad k_t \in [-\pi,\pi) \right\}.
\ee
For the sake of convenience, we introduce the time-dependent length
\be
l(t) \equiv \left| \frac{2}{h}\sin\left( \frac{ht}{2} \right)\right|,
\label{eq:loc-length}
\ee
which, as we shall see below, characterizes the dynamics of the system and is responsible for its localization for $h\neq 0$ within a typical distance
\be
\label{eq:l-loc}
l_{\rm loc} = \underset{t}{{\rm max}}\  l(t) = 2/|h|.
\ee
It is easy to show that, for any given value of $x$ such that $|x| \leq  l(t)$, one can determine two generically distinct values $k_F^-(x,t)$ and $k_F^+(x,t) \ge k_F^-(x,t)$ of $k$ on the Fermi contour at time $t$ corresponding to $x$, given by
\be
\begin{cases}
k_F^-(x,t) = ht/2 - \phi(t) + \arcsin\left(x/l(t)\right),\\
    k_F^+(x,t) =  ht/2 - \phi(t) + \pi - \arcsin\left(x/l(t) \right),
\end{cases}
\label{eq:kFpm}
\ee
where the phase $\phi(t)$ is defined such that 
\be
\phi(t) = 
\begin{cases}
0 & \mbox{for} \quad \sin(ht/2)>0,\\
\pi & \mbox{for} \quad \sin(ht/2)<0.
\end{cases}
\label{eq:def-phi}
\ee
The expressions in Eq.~\eqref{eq:kFpm} follow from inverting Eq.~\eqref{eq:Fermi_prof} which defines  the Fermi surface, suitably rewritten in the form $x_t = l(t)\sin(k_t -h t/2 + \phi(t))$.
Note that these values $k_F^\pm(x,t)$ play the role of local Fermi points, as explained in Ref.~\cite{rcdd-21}, and they have been carefully chosen here such that the vertical line $(x,k)$ in phase space with $k \in [k_F^-(x,t),k_F^+(x,t)]$ belongs to the region with $n(x,k;t)=1$.
For $|x| > l(t)$, instead, there are no such solutions $k_F^\pm(x,t)$, and for $x < l(t)$ or $x>l(t)$ the system behaves locally as a completely filled or empty Fermi sea, respectively.
The construction above is illustrated in Fig.~\ref{fig:Phase_space}, which provides a plot of the local occupation in phase space.

As a first application of this approach, we determine the particle density $\rho(x,t)$ and the current $j(x,t)$ from Eqs.~\eqref{eq:density} and \eqref{eq:current}, respectively. For $|x|\leq l(t)$ 
we get
\begin{align}
\rho(x,t) &= \int^{k^{+}_F(x,t)}_{k^{-}_F(x,t)} \frac{dk}{2\pi}  = \frac{k^{+}_F(x,t) - k^{-}_F(x,t)}{2\pi} = \frac{1}{\pi}\arccos\l \frac{x}{l(t)} \r, \label{eq:exp-rho}  \\
j(x,t) &=\int^{k^{+}_F(x,t)}_{k^{-}_F(x,t)} \frac{dk}{2\pi}  \sin k  = \frac{\cos{k_F^-(x,t)} - \cos{k_F^+(x,t)}}{2\pi} = \frac{1}{\pi}\sqrt{1-\frac{x^2}{l^2(t)}} \, \cos\left( \frac{ht}{2} - \phi(t)\right) \nonumber\\
&\phantom{=\int^{k^{+}_F(x,t)}_{k^{-}_F(x,t)} \frac{dk}{2\pi}  \sin k  = \frac{\cos{k_F^-(x,t)} - \cos{k_F^+(x,t)}}{2\pi} }= \frac{1}{2\pi}\sqrt{1-\frac{x^2}{l^2(t)}} \, \frac{\sin (ht)}{|\sin (ht/2)|}.
\label{eq:exp-j}
\end{align}
Note that $j(x,t)$ displays a discontinuity for $t=t_k = k T$ (with $T$ given in Eq.~\eqref{eq:expT}) because $j(x,t \to t_k^-) = - j(x,t \to t_k^+)$, i.e., the (non-vanishing) current changes direction at the beginning of each period of oscillation.
For $|x| > l(t)$, instead, the system does not evolve locally and a straightforward computation gives
\be
\rho(x,t) = \begin{cases} 1 \quad\mbox{for}\quad x<-l(t), \\ 0 \quad\mbox{for}\quad x>l(t),\end{cases} \quad\mbox{and}\quad j(x,t) = 0.
\label{eq:eqrho0}
\ee
Note that all the previous expressions for $\rho(x,t)$ and $j(x,t)$ are periodic in time with the period $T$ given by Eq.~\eqref{eq:expT} in spite of the fact that some intermediate steps of the calculation involve separately $\sin (ht/2)$ and $\cos(ht/2)$.

\begin{figure}[t]
    \centering
    \begin{subfigure}{0.49\textwidth}
         \centering
         \includegraphics[width=\linewidth]{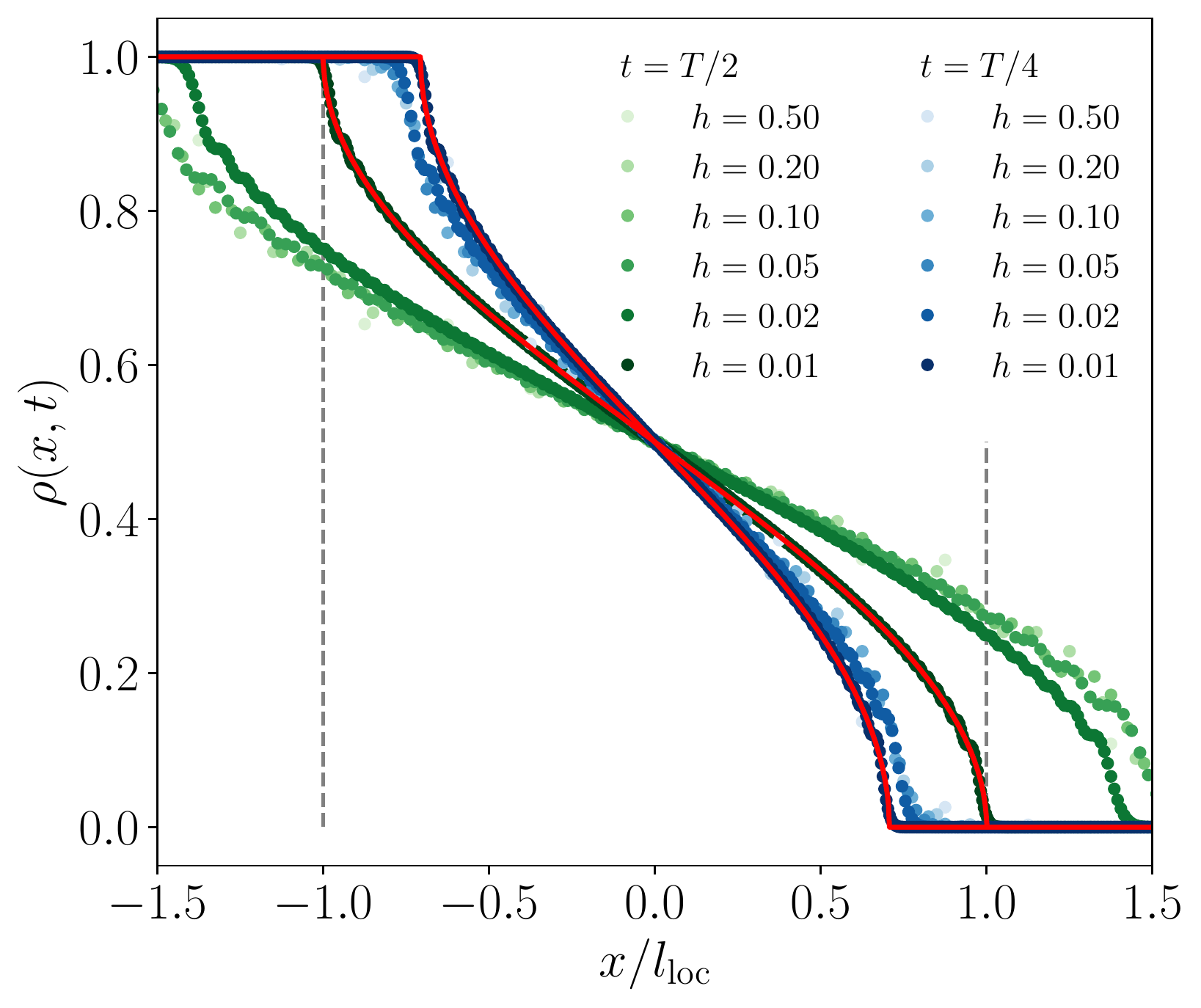}
         \label{fig:rho_profile}
     \end{subfigure}
     \begin{subfigure}{0.49\textwidth}
         \centering
         \includegraphics[width=\linewidth]{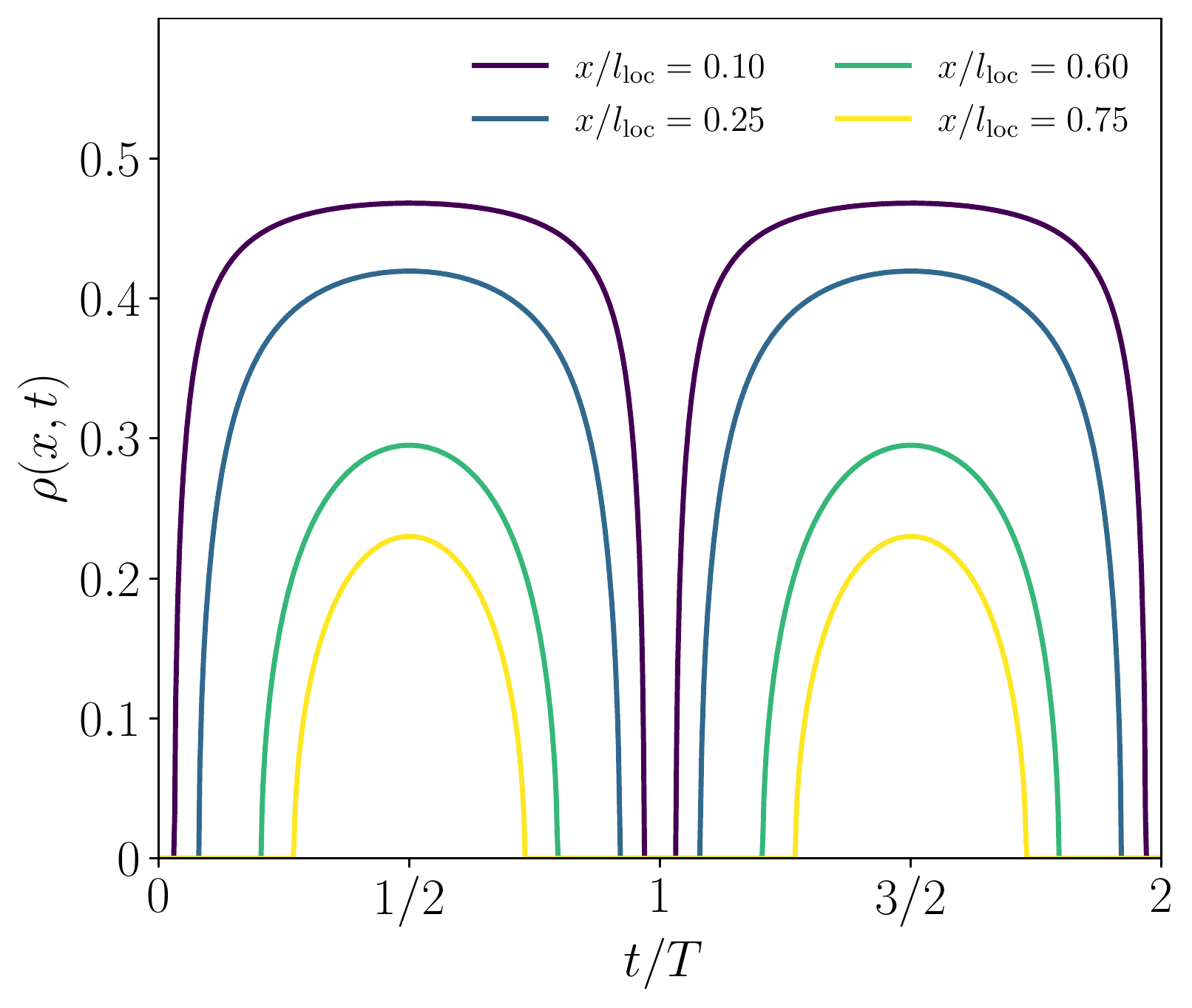}
         \label{fig:rho_time}
     \end{subfigure}
    \caption{Particle density $\rho(x,t)$ as a function of (left) $x$ for $t=T/2$ and $t=T/4$ or (right) $t$ for various values of $x$ along the chain. In both panels, the solid lines correspond to the GHD prediction in Eq.~\eqref{eq:exp-rho}.
    The dots in the left panel are the values obtained from the numerical computations with the lattice Hamiltonian, on a chain of length $L=600$ sites.
    By plotting these numerical data as functions of $x/l_{\rm loc}$ with $l_{\rm loc}$ given in Eq.~\eqref{eq:l-loc} one observes their convergence towards the GHD prediction as the value $h$ of the field decreases towards zero. 
    Note, however, that there are rather large deviations from the GHD prediction for large values of $h$, which are anyhow expected due to the highly fluctuating Bessel functions involved in the exact analytic prediction on the lattice, especially for $t\simeq T/2$. 
    As expected, for small values of $h$, the dynamics occurs only within the region $|x|<l_{\rm loc}$, delimited by the vertical dashed lines. 
    In the right panel, $\rho(x,t)$ displays the periodicity due to the Stark localization, with $\rho(x,t)=0$ at all times if $x/l_{\rm loc}>1$ or $\rho(x,t) = 1$ if $x/l_{\rm loc}<-1$.
    }
    \label{fig:rho}
\end{figure}

\begin{figure}[t]
    \centering
    \begin{subfigure}{0.49\textwidth}
         \centering
         \includegraphics[width=\linewidth]{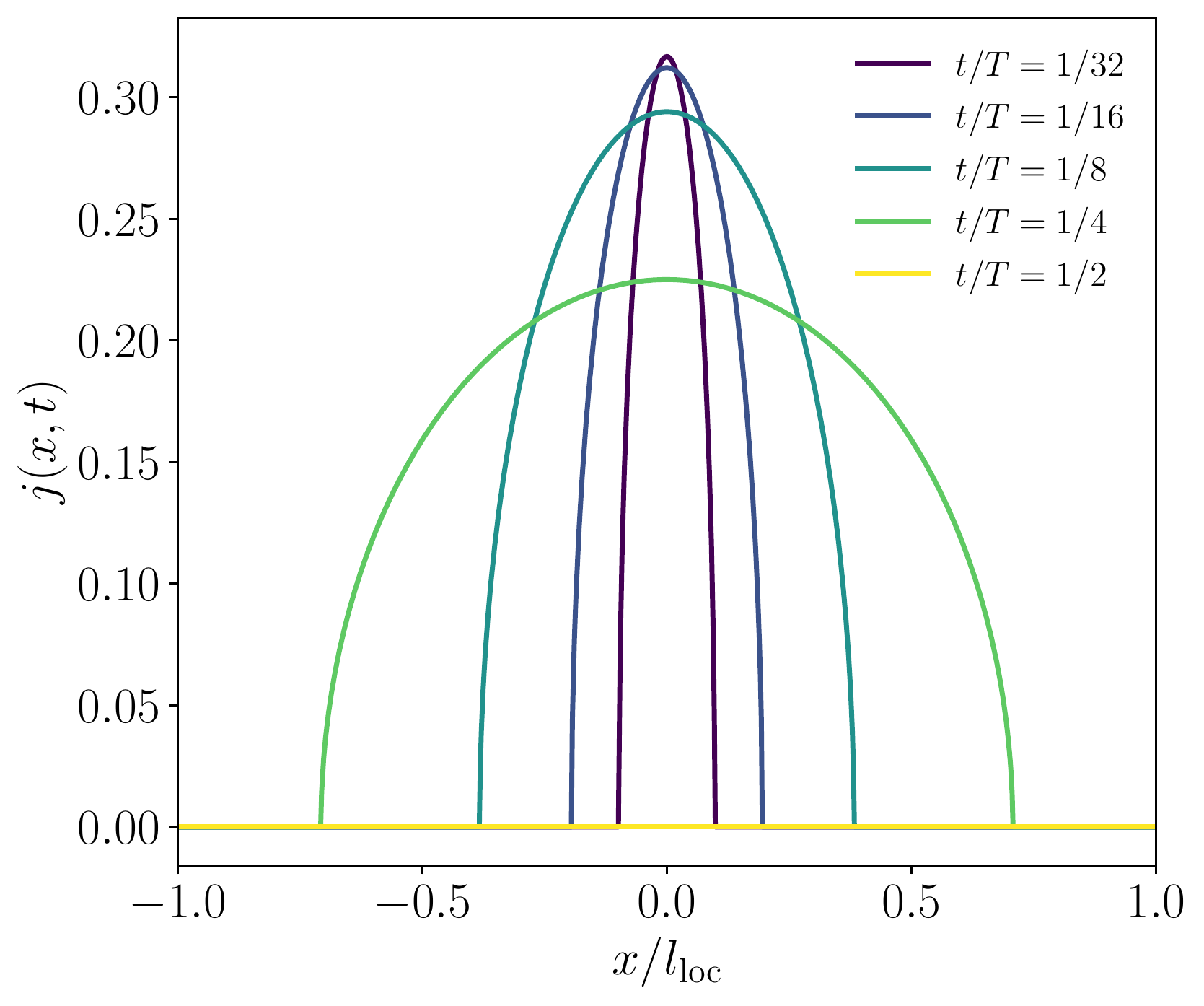}
         \label{fig:current_profile}
     \end{subfigure}
     \begin{subfigure}{0.49\textwidth}
         \centering
         \includegraphics[width=\linewidth]{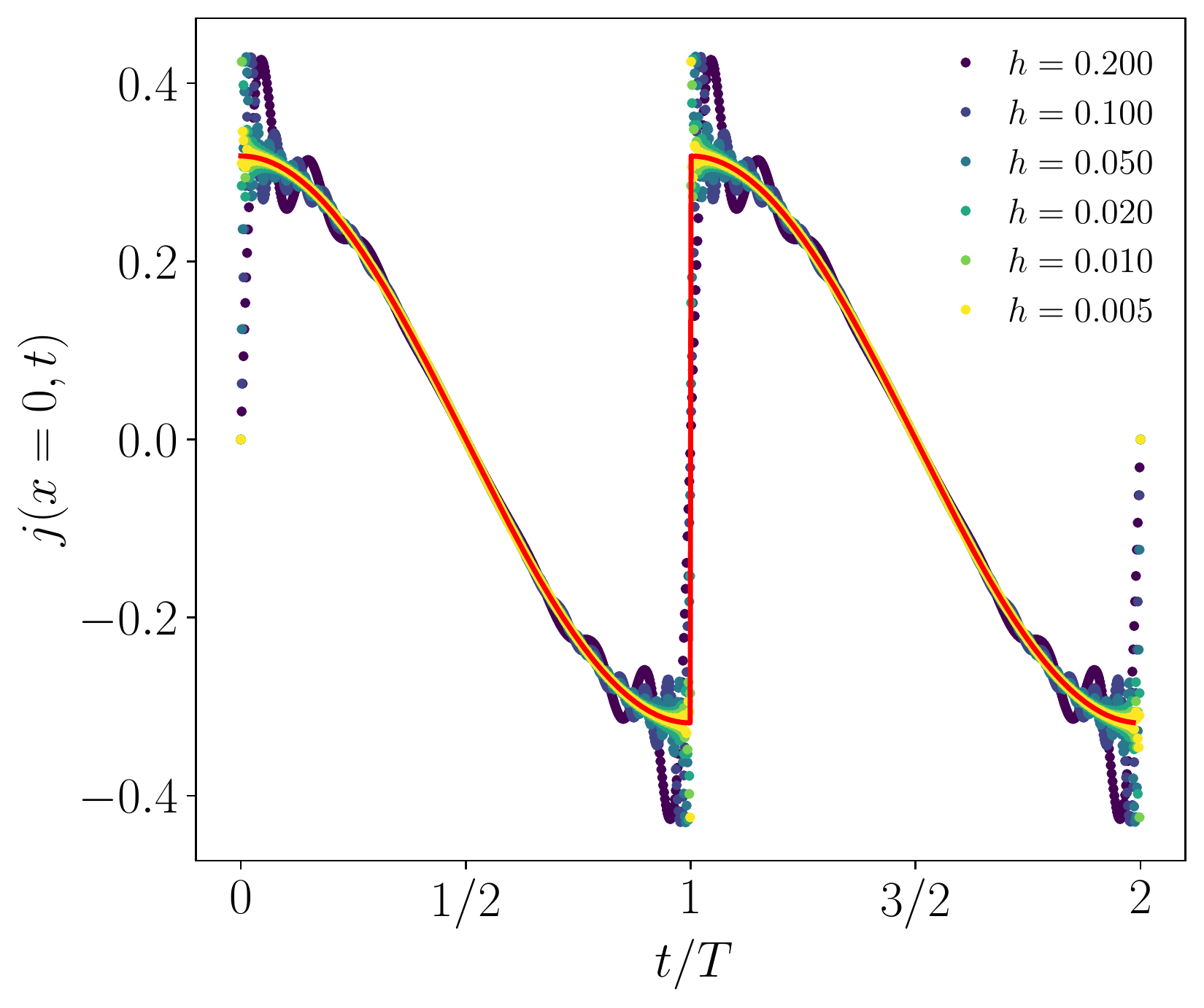}
         \label{fig:current_time}
     \end{subfigure}
    \caption{
    Particle current $j(x,t)$ as a function of (left) $x$ for various values of $t/T$ within a half-period of oscillation or (right) $t$ for $x=0$ and various values of $h$. In both panels, the solid lines correspond to the GHD prediction in Eq.~\eqref{eq:exp-j}.
    The coordinates $x$ and $t$ are rescaled by the natural $h$-dependent scales $l_{\rm loc}$ and $T$ [see Eqs.~\eqref{eq:loc-length} and \eqref{eq:expT}], respectively, so that their shapes in the hydrodynamic limit are actually independent of $h$. 
    In the right panel, symbols correspond to the values derived from the exact solution of the lattice model (see, c.f.,  Eq.~\eqref{eq:current_lattice_limit}). Upon reducing the strength of the external field $h$ and upon rescaling $t \to t/T$, the results on the lattice approach the curve predicted by GHD.}
    \label{fig:current}
\end{figure}

In Figs.~\ref{fig:rho} and \ref{fig:current} we plot the curves corresponding to Eqs.~\eqref{eq:exp-rho}  and \eqref{eq:exp-j}, respectively, and we compare them with the result of numerical calculations on the lattice in the hydrodynamic regime.
The numerical data for the density profile have been obtained by computing the time evolution of the correlation matrix in Eq.~\eqref{eq:Cmatrix} and then by considering its diagonal elements, as explained in Appendix~\ref{sec:app:numerics}. For the current, instead, we used the analytical result on the lattice reported in, c.f., Eq.~\eqref{eq:current_lattice_limit} of Appendix~\ref{sec:app:lattice}.

The expressions derived above can be used also to investigate the limit $h\rightarrow 0$, --- corresponding to the melting of a domain wall in a homogeneous chain, --- which was studied in Refs.~\cite{skcd-21,arrs-99,scd-22,pk-07}. 
In that case, $l(t) = t$ [see Eq.~\eqref{eq:loc-length}], the oscillations disappear, and Eq.~\eqref{eq:exp-j} for $|x|<t$ gives
\begin{align}\label{eq:rho_J_h0} 
\begin{cases}
\rho(x,t)= (2/\pi)\arccos\l x/t \r, \\
j(x,t)= (1/\pi)\sqrt{1-x^2/t^2},
\end{cases}
\end{align}
which coincides with the results of Ref.~\cite{scd-22} for a single domain wall.
As pointed out in Refs.~\cite{bglsv-22,bglsv-23}, it is worth noticing that, for a given $x$, the value of $\rho(x,t)$ at $h\neq 0$ [see Eq.~\eqref{eq:exp-rho}] can be easily obtained from its value at $h=0$ in Eq.~\eqref{eq:rho_J_h0}, via the substitution
\be
t\rightarrow l(t).
\label{eq:t-to-l}
\ee
However, this does not hold for the current $j(x,t)$.
To better understand the origin of these facts, it is sufficient to compare the Fermi contour for $h\neq 0$ with that for $h=0$. In particular, we observe that the former, given by 
\be
\left\{ (x,k) | \ x = l(t) \sin \l k-\frac{h t}{2} + \phi(t)\r \quad\mbox{and} \quad k \in [-\pi,\pi)\right\},
\ee
[see the parameterization of Eq.~\eqref{eq:Fermi_prof} introduced after  the definition of $\phi(t)$ in Eq.~\eqref{eq:def-phi}] is recovered from the latter, i.e., 
\be
\left\{ (x,k) | \  x = t \sin k \quad\mbox{and} \quad k \in [-\pi,\pi) \right\},
\ee
via a reparameterization of time $t \rightarrow l(t)$, followed by a shift of the momentum $k\rightarrow k-h t/2 + \phi(t)$.
Since the density $\rho(x,t)$ does not depend on momenta [see its semi-classical expression in Eq.~\eqref{eq:density}], it is not sensitive to such a shift, and therefore the overall effect on $\rho(x,t)$ of having a linear potential simply amounts at a reparameterization of time, as observed in Refs.~\cite{bglsv-22,bglsv-23}.
However, this does not apply to the current $j(x,t)$ because its expression in Eq.~\eqref{eq:current} is not invariant under such a momentum shift.

\subsection{Quantum GHD: Entanglement entropy and full counting statistics}

We proceed further with the analysis of the domain-wall dynamics, and we aim at characterizing the entanglement among complementary spatial regions. While the entanglement in the presence of Stark localization has been studied, so far, numerically~\cite{eip-09,bs-18}, by using CFT in curved space-time~\cite{eb-17,dsvc-17,adsv-16} or by exploiting the substitution $t \to l(t)$~\cite{bglsv-23}, here we derive analytically its dynamics on the basis of a quantized version of GHD \cite{rcdd-20,rcdd-21}.

Let us consider a bipartition $A \cup \bar{A}$ of the lattice in two extended and complementary subsystems $A$ and $\bar{A}$. Given the reduced density matrix of $A$
\be
\rho_A(t) \equiv \text{Tr}_{\bar{A}}\l e^{-iHt}\ket{\Psi_0}\bra{\Psi_0}e^{iHt}\r,
\ee
a good entanglement measure between $A$ and $\bar{A}$ is provided by the von Neumann entropy (also known as entanglement entropy), given by
\be
S(t) \equiv - \text{Tr}\l \rho_A(t)\log \rho_A(t)\r.
\ee
Being this quantity highly non-local in space, one may ask whether it is possible to determine it via the local description provided by hydrodynamics.
It turns out that, for states with short-range correlations, the semi-classical Yang-Yang entropy \cite{yy-69} is able to capture the leading contribution to the entanglement entropy \cite{bfpc-18}, and therefore one gets
\be
S(t) \simeq -\int_{-\pi}^\pi\! \frac{dk}{2\pi} \int_A dx \left[ n \log n + (1-n)\log (1-n) \right]_{n = n(x,k;t)}.
\ee
However, the local occupation number $n(x,k;t)$ is either $0$ or $1$ in the system under consideration here and therefore the semi-classical expression above vanishes, while the entanglement entropy does not. 
A solution to this apparent contradiction has been put forward in Refs.~\cite{skcd-21,rcdd-21,dsvc-17} by generalizing the standard GHD to what has been dubbed quantum GHD, which accounts for quantum effects beyond the semi-classical approximation. Indeed, while the Yang-Yang entropy predicts the extensive contribution to $S(t)$, which vanishes, the dominant sub-extensive contribution is correctly predicted by quantum GHD.

\begin{figure}
    \centering
    \begin{subfigure}{0.49\textwidth}
         \centering
         \includegraphics[width=\linewidth]{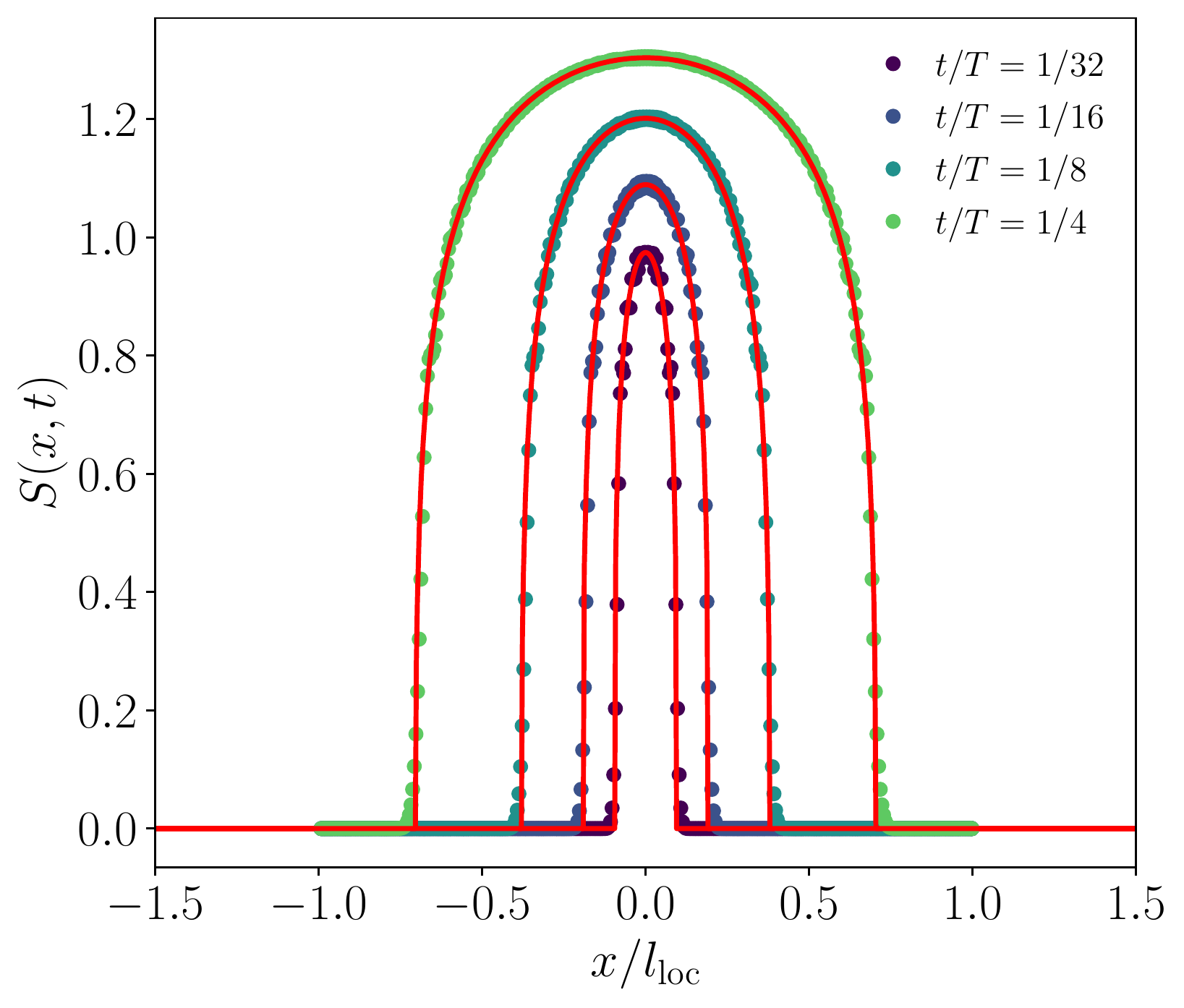}
         \label{fig:entanglement_profile}
     \end{subfigure}
     \begin{subfigure}{0.49\textwidth}
         \centering
         \includegraphics[width=\linewidth]{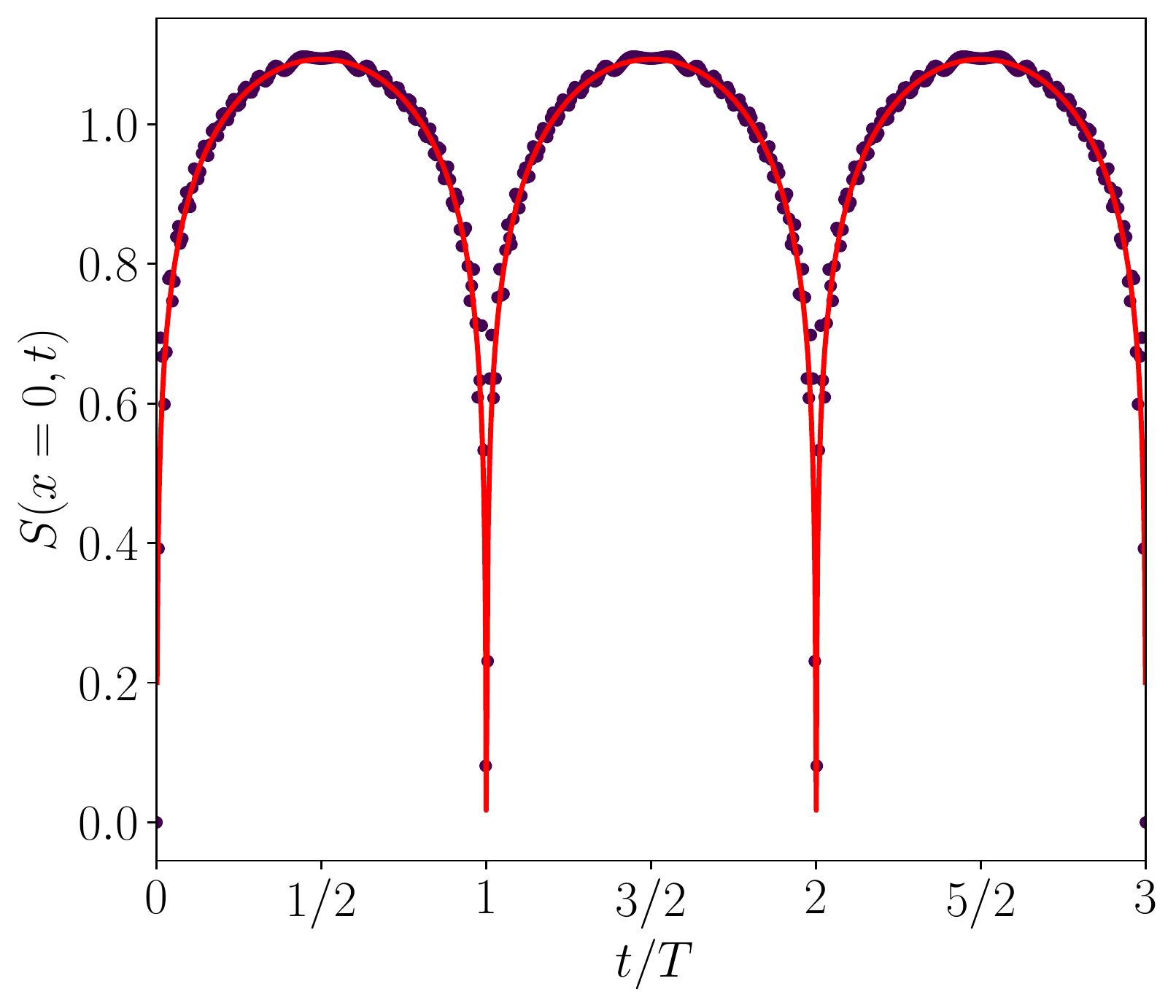}
         \label{fig:entanglement_time}
     \end{subfigure}
    \caption{Evolution of the von Neumann entanglement entropy $S(x,t)$ between $A=[x, \infty)$ and $\bar{A}$, starting from a single domain wall located at $x=0$. In the left panel $S(x,t)$ is plotted as a function of $x$ for various values of time $t$, while the right panel shows the periodic evolution of $S(x=0,t)$ as a function of $t$. As in the previous figures, $x$ and $t$ are rescaled by the natural $h$-dependent scales $l_{\rm loc}$ and $T$ [see Eqs.~\eqref{eq:loc-length} and \eqref{eq:expT}], respectively. In both panels, solid lines correspond to the prediction of quantum GHD in Eq.~\eqref{eq:entanglement_vN}, while symbols indicate the  numerical results obtained for a chain of (left) $L=400$ sites with field $h=0.01$ or (right) $L=300$ sites with $h=0.05$.}
    \label{fig:entanglement}
\end{figure}

This approach is well established for the domain-wall state in the absence of the external potential (i.e., for $h=0$), where the dynamics of the entanglement entropy, as well as other entanglement measures (e.g., R\'enyi entropies, full counting statistics, charged moments), have been studied~\cite{scd-22,sh-22,asw-22}.
Our goal here is to generalize that method in the presence of a linear potential, thus characterizing analytically the Bloch oscillation of the entanglement entropy.
We anticipate here that, as observed in Ref.~\cite{bglsv-23}, the evolution of $S(t)$ for $h \neq 0$ can be recovered from the one at $h = 0$ via the substitution in Eq.~\eqref{eq:t-to-l}, as discussed in Sec.~\ref{sec:GHD_Stark}.
While this might be expected, as the measures of spatial entanglement considered here should be insensitive to momentum shifts, it is actually a non-trivial fact because our analysis based on quantum GHD goes beyond the semi-classical description to which the previous heuristic argument actually applies.
Following Refs.~\cite{cc-04,cc-09,ccd-08}, 
we employ the  replica trick, and we first compute the $n$-th R\'enyi entropy
\be\label{eq:Renyi_def}
S_n(t) \equiv \frac{1}{1-n}\log \text{Tr}\l \rho_A^n(t)\r,
\ee
for integer $n\geq 2$, and we eventually continue the results to $n=1$ in order to get the entanglement entropy
\be
S(t) = \underset{n \rightarrow 1}{\lim}S_n(t).
\ee
For the sake of simplicity, we focus here on the case $A = [x,\infty)$, i.e., on the half-chain starting from $x$. Then we express $\text{Tr}\l \rho^n_A(t)\r$ in terms of the expectation value of a twist field $\mathcal{T}_n(x)$~\cite{ccad-07,cc-09}, which acts as a cyclic permutation over the region $A$, as
\be
\text{Tr}\l \rho^n_A(t)\r = \la \mathcal{T}_n(x,t)\ra  \equiv {}^n\!\bra{\Psi_0}e^{iHt}\mathcal{T}_n(x)e^{-iHt}\ket{\Psi_0}^{n},
\ee
where $\ket{\Psi_0}^{n}$ denotes the replicated initial state.
We explain below the quantum GHD, following closely Ref. \cite{scd-22}, which gives $\la \mathcal{T}_n(x,t)\ra$ in terms of a chiral conformal field theory (CFT) associated to the Fermi contour in phase space. We focus on a partition with $|x|\leq l(t)$, thus having non-trivial dynamics and such that two corresponding Fermi points $k^{\pm}_F(x,t)$ are present.
We parameterize the Fermi contour through an angular variable $\theta \in [-\pi,\pi]$ and we decompose $\mathcal{T}_n(x,t)$ in a pair of chiral twist fields in the CFT, denoted by $\tau_n(\theta^{+})$ and $\tilde{\tau}_n(\theta^{-})$, with $\theta^{\pm}$ corresponding to the Fermi points $(x,k_F^{\pm}(x,t))$~\cite{scd-22}.
For the sake of convenience, we identify $\theta$ as the momentum $k$ corresponding to a generic point $(x,k)$ of the Fermi contour, and we set $\theta^\pm = k^\pm_F$. 
Eventually, one expresses the expectation value of the twist field as \cite{scd-22}
\be\label{eq:QGHD_TField}
\la \mathcal{T}_n(x,t)\ra = \l \varepsilon_n(x,t)\r^{2h_n} \la \tau_n(\theta^{+}) \tilde{\tau}_n(\theta^{-})\ra \left| \frac{d\theta^+}{dx} \right|^{h_n} \left|\frac{d\theta^-}{dx}\right|^{h_n},
\ee
where $h_n = \l n-n^{-1}\r/24$ is the conformal dimension of $\tau_n$, $\varepsilon_n(x,t)$ is given by~\cite{dsvc-17,jk-04,cc-05,ce-10}
\be
\label{eq:epsilon}
\varepsilon_n(x,t) = \frac{\varepsilon_n}{\sin\l \pi \rho(x,t)\r},
\ee
and $\varepsilon_n$ is a non-universal UV cutoff. We compute the two point-function
\be
\label{eq:twopoint_twist}
\la \tau_n(\theta^{+}) \tilde{\tau}_n(\theta^{-})\ra \equiv \left[ \frac{1}{ 2\sin \l \frac{\theta^+-\theta^-}{2}\r}\right]^{2h_n},
\ee
fixed by conformal invariance, we express the Jacobian as
\be
\left|\frac{d\theta^\pm}{dx}\right| = \left|\frac{dk^{\pm}_F(x,t)}{dx}\right| = \frac{1}{l(t)\sqrt{1 - x^2/l^2(t)}},
\label{eq:exp-Jac}
\ee
and, from Eq.~\eqref{eq:QGHD_TField}, we eventually get
\be
\la \mathcal{T}_n(x,t)\ra = \left[ \frac{\varepsilon_n}{2 l(t)}\l 1- \frac{x^2}{l^2(t)} \r^{-3/2}\right]^{2h_n}.
\ee
Inserting this expression in Eq.~\eqref{eq:Renyi_def}, we determine the analytic form of the R\'enyi entropies
\be
\label{eq:ent_Stark}
S_n(t) = \frac{2h_n}{n-1}\log \l l(t)\l 1- \frac{x^2}{l^2(t)} \r^{3/2} \r + \dots,
\ee
up to an omitted, non-universal constant. We note that this kind of calculation for $h=0$ can be found in Ref.~\cite{scd-22}; the only difference compared to the present analysis is the expression of $\rho(x,t)$ and $k^{\pm}_F(x,t)$.
As anticipated above, the parameter $h$ enters in Eq.~\eqref{eq:ent_Stark} only via $l(t)$, which, as anticipated, amounts at replacing $t$ with $l(t)$ [see Eq.~\eqref{eq:t-to-l}] in the prediction of Ref.~\cite{dsvc-17}.
Finally, by taking the analytic continuation $n\rightarrow 1$, we get the von Neumann entropy
\be
\label{eq:entanglement_vN}
S(t) = \frac{1}{6}\log \l l(t)\l 1- \frac{x^2}{l^2(t)} \r^{3/2} \r + \gamma,
\ee
where the non-universal constant $\gamma \simeq 0.4785\dots$ is extracted from the result of Ref.~\cite{jk-04}.
In Fig.~\ref{fig:entanglement} we plot the quantum GHD prediction for the entanglement entropy compared with numerical data on the lattice in the hydrodynamic regime, finding perfect agreement.
Beyond the entanglement entropy, the approach discussed above allows us to characterize also the fluctuations of the number of fermions, as explained in Ref.~\cite{sh-22}, and to predict analytically the full counting statistics. In fact, consider the operator
\be
N_A \equiv \sum_{x \in A} \psi^\dagger_x\psi_x,
\ee
i.e., the number of fermions in the spatial region $A$.
Its expectation value at time $t$ is predicted by GHD to be given by the following semi-classical expression: 
\be\label{eq:N_average}
\la N_A(t)\ra \simeq \int_{x \in A}\!\!dx \ \rho(x,t).
\ee
Although all higher-order connected moments (which describe the quantum fluctuations related to the entanglement between $A$ and $\bar{A}$) vanish at the semi-classical level, their leading behavior can be computed via quantum GHD. In order to do this, we focus on the full-counting statistics 
\be
\la e^{i\alpha N_A(t)}\ra,
\ee
of $A$, i.e., the generating function of the moments of $N_A$.
As done above and for the sake of simplicity, we consider $A = [x,+\infty)$ and follow the same construction as that previously illustrated for the R\'enyi entropies.
Here, it is important to identify the fields in the chiral CFT corresponding to $e^{i\alpha N_A}$, which are (chiral) vertex $U(1)$ fields $V_{\pm \alpha}$.  Their conformal dimension is given by
\be
\label{eq:conf_dimension}
h_{\alpha} = \frac{1}{2}\l \frac{\alpha}{2\pi}\r^2, \quad\mbox{with}\quad \alpha \in [-\pi,\pi].
\ee
Eventually, it is possible to express \cite{sh-22}
\be
\frac{\la e^{i\alpha N_A(t)}\ra}{e^{i\alpha \la N_A(t)\ra }} = \left[ \varepsilon_\alpha(x,t)\right]^{2h_\alpha} \la V_{+\alpha}(\theta^{+}) V_{-\alpha}(\theta^{-})\ra \left| \frac{d\theta^+}{dx} \right|^{h_\alpha} \left|\frac{d\theta^-}{dx}\right|^{h_\alpha},
\label{eq:CFT-count}
\ee
and therefore, by using Eqs.~\eqref{eq:exp-Jac}, \eqref{eq:epsilon}, and \eqref{eq:twopoint_twist},
\be
\label{eq:full_counting_res}
\log \la e^{i\alpha N_A(t)}\ra = i\alpha \la N_A(t)\ra -  \l \frac{\alpha}{2\pi}\r^2\log \l l(t) \l 1- \frac{x^2}{l^2(t)} \r^{3/2} \r + \l \frac{\alpha}{2\pi}\r^2 \log \l \frac{\varepsilon_\alpha}{2} \r ,
\ee
with $\varepsilon_\alpha$ being a $\alpha$-dependent non-universal UV-cutoff.
Since the dependence on the CFT fields of Eqs.~\eqref{eq:QGHD_TField} and \eqref{eq:CFT-count} enters only via the scaling dimensions of the involved fields, it is sufficient to replace $h_n\rightarrow h_\alpha$ in Eq.~\eqref{eq:QGHD_TField} in order to get $\la e^{i\alpha N_A(t)}\ra$.
We emphasize that while the average $\la N_A(t)\ra$ is not directly predicted by field theory, it can be computed by GHD via Eqs.~\eqref{eq:N_average} and \eqref{eq:exp-rho}, and for $A=[0,\infty)$ it is given by
\be
\label{eq:average_FCS}
\la N_{A=[0,\infty)}(t)\ra = \frac{l(t)}{\pi}.
\ee
Finally, it is worth mentioning that in the large-scale limit, namely under $t \rightarrow \lambda t$, $x \rightarrow \lambda x$, $h \rightarrow h/\lambda$, being $\lambda$ a large dimensionless parameter, the average number of particles scales extensively as $\la N_A(t)\ra \rightarrow \lambda\la N_A(t)\ra$, while its variance grows logarithmically as $\la N^2_A(t)\ra_c \rightarrow \log(\lambda) \, \la N^2_A(t)\ra_c$, where $\langle\cdots\rangle_c$ stands for cumulants. By contrast, higher-order cumulants, which appear in Eq.~\eqref{eq:full_counting_res} due to the $\alpha$-dependent cutoff as powers of $\alpha$ larger than two, are finite as $\lambda\rightarrow \infty$, but cannot be determined within the quantum GHD formalism.
These are typical features of free fermions at equilibrium~\cite{kl-09,cmv-12}, which might be affected, e.g., by the presence of defects~\cite{cmc-23}. In our case, these properties can be traced back to the fact that the scaling dimension of the $U(1)$ vertex fields $V_{\pm}(\alpha)$ is proportional to $\alpha^2$, see Eq.~\eqref{eq:conf_dimension}.

\section{Stark localization in a generic potential}
\label{sec:Stark_generic}

\begin{figure}
    \centering
    \includegraphics[width=0.6\textwidth]{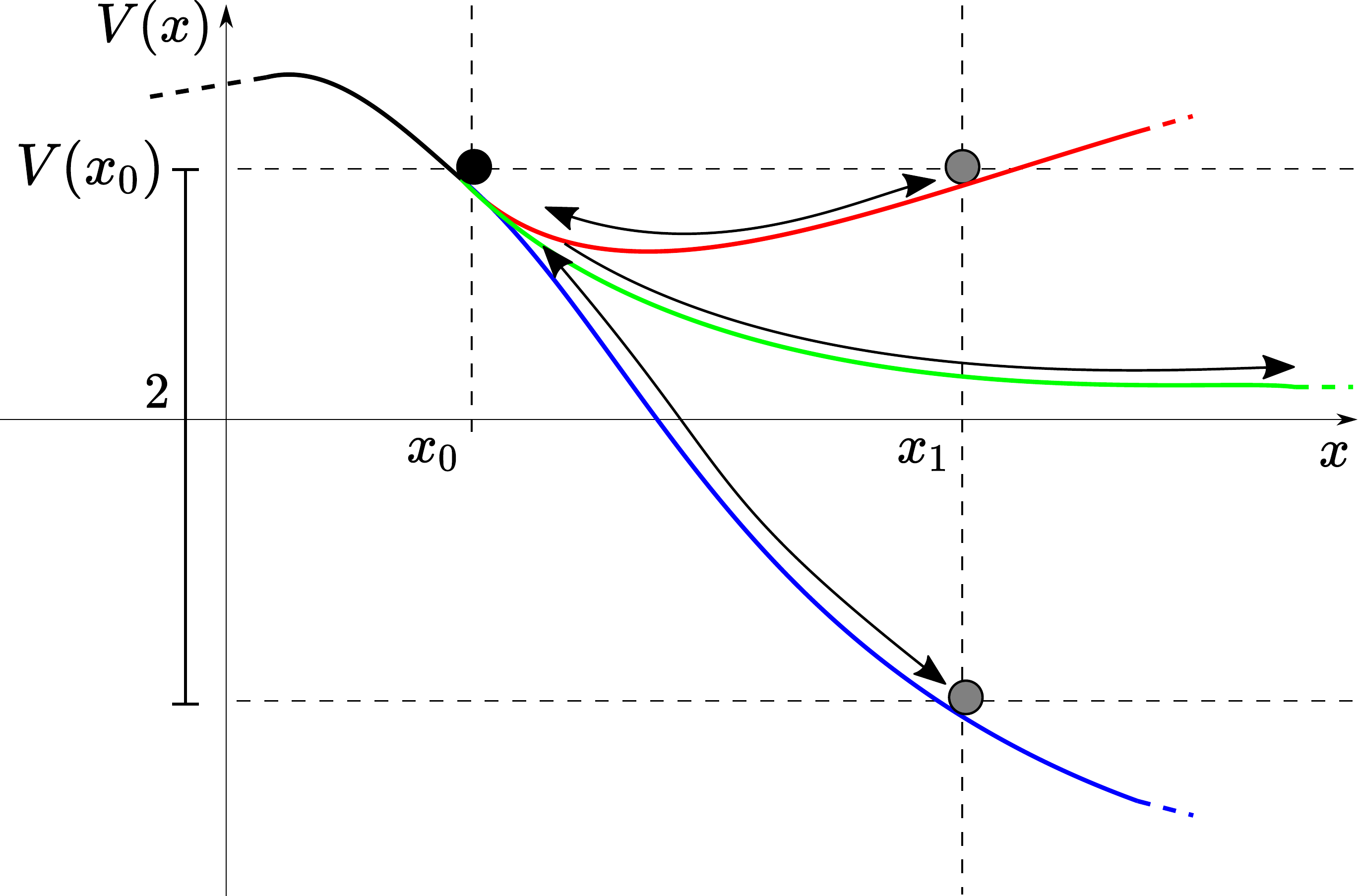}
    \caption{Different kinds of dynamics in a generic potential, in the presence of a bounded kinetic energy, see Eq.~\eqref{eq:Ham_V}. The black particle at $x=x_0$ has zero momentum at $t=0$ and the curves with different colors represent various possible potentials.
    If the particle is subject to the potential given by the red curve, it has a periodic motion, oscillating between $x_0$ and $x_1$, where $x_1$ is such that  $V(x_0) = V(x_1)$. In the blue potential, the particle oscillates between $x_0$ and $x_1$ with $V(x_0) - V(x_1) = 2$ because of Stark localization. In the case of the green curve, instead, the particle moves forever, since there is no $x_1 > x_0$ such that $V(x_0) - V(x_1) = 2$.}
    \label{fig:potential_Stark}
\end{figure}

In this section, we go beyond the analysis of the linear potential, and we study the semi-classical dynamics of the Hamiltonian \eqref{eq:Ham_V}. 
We first provide an argument to establish the conditions under which a trajectory starting from $(x,k)=(x_0,0)$ at time $t=0$, having an initial vanishing velocity $v_0 = v(k=0)=0$, experiences Stark localization in the presence of a generic potential $V(x)$. This is a relevant question for a wider class of systems, e.g., the long-range interacting model which, in this respect, was investigated in Ref.~\cite{lzsg-19}.
We anticipate here that the analysis presented below readily extends to the somehow equivalent initial condition in which the particle has, as above, a vanishing initial velocity $v_0 = v(k_0)=0$ but with a non-vanishing wave-vector $k_0=\pm \pi$. 
In this case, the subsequent dynamics of the particle starting from $x_0$ occurs in the direction in which the potential $V(x)$ increases, because the value of the kinetic term in Eq.~\eqref{eq:Ham_V} can only decrease compared to its initial (maximum) value.
Without loss of generality, we assume 
\be
V'(x_0) <0,
\ee
as the analysis for $V'(x_0)>0$ would be identical, while if $V'(x_0)=0$ the trajectory reduces just to the initial point as no evolution occurs within the semi-classical approximation.
Under the above assumption, the particle starting at $(x,k)=(x_0,0)$ is initially accelerated to the right of $x_0$. Then, it either stops at a certain position $x=x_1>x_0$ or it escapes towards infinity.
If it stops, its velocity $v(k)=\sin k$ at  $x=x_1$ has to vanish and therefore the corresponding momentum $k_1$ of the particle is either $k_1=0$ or $k_1=\pi \, (\,=-\pi)$.
By conservation of energy, one easily shows that in the former case 
\be
V(x_0) - V(x_1) = 0,
\label{eq:cond-per}
\ee
which corresponds to usual oscillations around the local \emph{minimum} of the confining potential, while 
\be
V(x_0)-V(x_1)=2
\label{eq:cond-Stark}
\ee
in the latter. 
This means that if the potential $V(x)$ for $x>x_0$ is bounded by
\be
V(x)\in (V(x_0)-2,V(x_0))
\label{eq:cond-escape}
\ee
the particle cannot actually stop and reverse the direction of its motion and thus it moves towards $x\to+\infty$. 
Conversely, if this is not the case, one can identify the smallest value $x_1$ of $x$ for which $V(x)$ escapes the interval in Eq.~\eqref{eq:cond-escape} from below. The existence of this $x_1$ implies that the dynamics of the particle occurs within the region $x \in [x_0,x_1]$. 
Accordingly, $V(x_0)-V(x_1)$ being equal to 0 or 2 indicates either usual periodic motion or Stark localization, respectively.
The various cases discussed above are illustrated in Fig.~\ref{fig:potential_Stark}: the black particle starting at $x=x_0$ undergoes usual oscillations if the potential is given by the red curve [Eq.~\eqref{eq:cond-per}], it experiences Stark localization if the potential is the one indicated in blue [Eq.~\eqref{eq:cond-Stark}], while it escapes to infinity in the potential given by the green curve [Eq.~\eqref{eq:cond-escape}].
If the starting point is $(x,k) = (x_0,\pm \pi)$, the same analysis as the one done above indicates that Stark localization occurs for $V(x_0) - V(x_1) = -2$, while if $V(x_0) - V(x_1) = 0$, the particle oscillates around the local \emph{maximum} of the potential occurring within the interval $[x_0,x_1]$.

\subsection{Topological properties of the Hamiltonian flow}

Here we adopt a topological perspective, and we analyze the way the trajectories of a particle foliate the entire phase space. This is particularly useful in this context because we argued above that Stark localization is connected to a topological property of the particle trajectories in phase space, i.e., their winding around the Brillouin zone.
Let us consider the isoenergetic surface $\Gamma_E$ with energy $E$, defined by
\be\label{eq:Energy_Set}
\Gamma_E = \{(x,k) | \mathcal{H}(x,k) = E\},
\ee
with $\mathcal{H}$ given by Eq.~\eqref{eq:Ham_V}.
In general, $\Gamma_E$ is the union of many disconnected trajectories, some of which might experience Stark localization. As the energy $E$ is varied, $\Gamma_E$ may change its topology due to the presence of critical points $(x_c,k_c)$ of the map $(x,k) \rightarrow \mathcal{H}(x,k)$. These points are defined by the condition
\be
{\bm \nabla} \mathcal{H}(x_c,k_c) ={\bm 0}, \quad \mbox{i.e.,} \quad \begin{cases} k_c=0 \quad\mbox{or}\quad  \pm\pi, \\ V'(x_c)=0.\end{cases}
\ee
In other words, whenever $V(x)$ has a local minimum or maximum, 
a pair of critical points might appear in the phase space depending on the value of $E$.
In order to characterize the nature of a possible critical point  $(x_c,k_c)$, we linearize the dynamics around it, assuming for the sake of simplicity that $V''(x_c)\neq 0$. We define the deviation from the stationary points as
\be
(\delta x,\delta k) \equiv (x-x_c, k-k_c),
\ee
and for $k_c=0$ we get
\be
\label{eq:LD-0}
\begin{pmatrix}\delta \dot{x} \\ \delta 
 \dot{k} \end{pmatrix} = \begin{pmatrix} 0 & 1 \\ -V''(x_c) & 0 \end{pmatrix}\begin{pmatrix}\delta x \\ \delta 
k \end{pmatrix} + {\cal O}(\delta x^2, \delta x\delta k, \delta k^2),
\ee
while for $k_c=\pm\pi$ we have
\be
\label{eq:LD-pi}
\begin{pmatrix}\delta \dot{x} \\ \delta 
 \dot{k} \end{pmatrix} = \begin{pmatrix} 0 & -1 \\ -V''(x_c) & 0 \end{pmatrix}\begin{pmatrix}\delta x \\ \delta 
k \end{pmatrix} + {\cal O}(\delta x^2, \delta x\delta k, \delta k^2).
\ee
Accordingly, from the sign of the determinants of the linearized maps above, one concludes that the critical point can be either elliptic (with $V''(x_c)>0$ and $k_c=0$ or $V''(x_c)<0$ and $k_c=\pi$) or hyperbolic (with $V''(x_c)<0$ and $k_c=0$ or $V''(x_c)>0$ and $k_c=\pi$). 

We now investigate
how the qualitative features of the possible periodic dynamics of the particle change when its energy $E$ approaches a critical value $E_c$, i.e., a value for which the corresponding isoenergetic surface contains the critical points identified above. 
Consider first the case of a Stark-localized trajectory of energy $E$, which encircles periodically the Brillouin zone in a finite time. According to the characterization of these trajectories discussed above (see Fig.~\ref{fig:potential_Stark}), if we slightly change the energy $E$, the resulting perturbed trajectories would generically be still Stark localized and periodic.
However, when $E$ approaches a critical value $E_c$, the corresponding trajectory of the particle gets close to a separatrix and, as a result, its period diverges. 
Upon crossing that critical value $E_c$, the trajectory might change its topology, ceasing to be Stark localized. The same conclusions apply to the other type of periodic orbits we are interested in, i.e., those corresponding to the usual periodic motion in which the trajectory does not encircle the first Brillouin zone. This means that a change of the qualitative features of the trajectories occurs only upon crossing a critical value of the energy.

It is then natural to ask whether it is possible to predict the topology of the trajectories within an interval of energy $E$ delimited by two consecutive critical values $E_{0,1}$, i.e., $E\in (E_0,E_1)$.
In this respect, we point out that a local analysis of the Hamiltonian flow at its stationary points is not sufficient in this respect, as the following paradigmatic example demonstrates. 
Consider, in fact, an unbounded potential $V(x)$ such that
\be
V(x\to \pm \infty) = \mp \infty,
\ee
with, say, a local minimum at $x=0$ and a local maximum at $x=1$. For simplicity, assume that $V(0)=0$ while we vary the value of $V(1)>0$. A possible instance of this potential provided by
\be
V(x) = - V(1) \, x^2(2x-3).
\label{eq:pot-ex}
\ee
The Hamiltonian $\mathcal{H}(x,k)$ with this potential has critical points with $x_c = 0$ or $1$ and $k=0$ or $\pi$ and corresponding critical energies $\{ -1, -1+V(1), 1, 1+V(1)\}$.
Clearly, the existence and location of these critical points and the hyperbolic/elliptic character of the corresponding linearized dynamics in Eqs.~\eqref{eq:LD-0} and \eqref{eq:LD-pi} are not affected by the actual value of $V(1)$.

\begin{figure}[t]
    \centering
    \includegraphics[width=\textwidth]{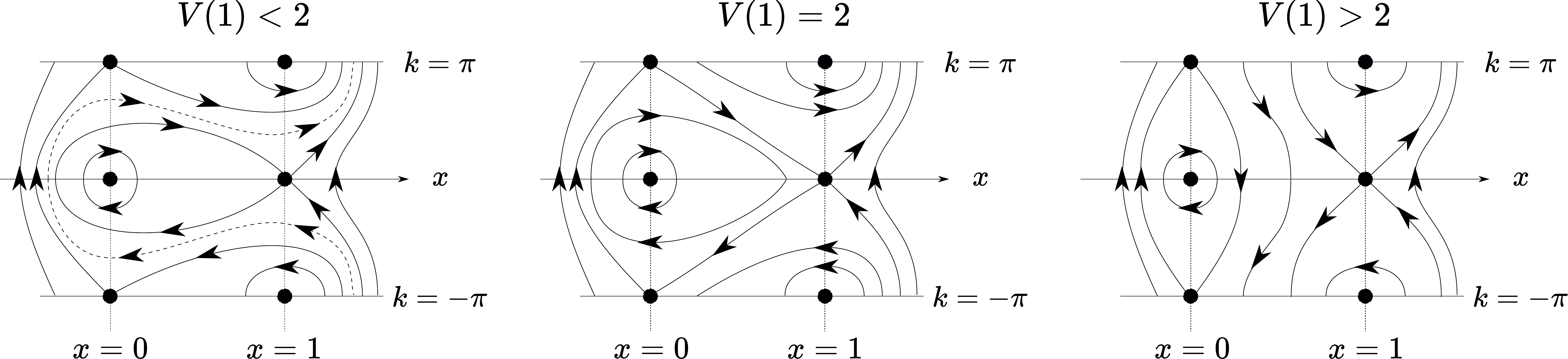}
    \caption{Illustration of the change in the topology of the trajectories when the value $V(1)>0$ of the local maximum of the potential in Eq.~\eqref{eq:pot-ex} at $x=1$ is varied. For $V(1)<2$ (left) there are trajectories which start at $(x<0, k=0)$ and visit positions with $x>1$, being the potential barrier sufficiently small. These trajectories do not experience Stark localization for $x \in [0,1]$ while they are localized only for $x>1$ (dashed line).
    For $V(1)=2$ (center) a separatrix which connects the two hyperbolic points at $x=0$ and $x=1$ appears, as a consequence of the degeneracy in the critical energies. When $V(1)>2$ (right) the two regions $x<0$ and $x>1$ are dynamically disconnected. This happens either because the high potential barrier confines the trajectory, or because Stark localization occurs.}
    \label{fig:phase_space_generic}
\end{figure}

However, a transition appears for $V(1)=2$, namely, if $V(1)>2$ there are some Stark localized trajectories with energy $E \in (-1,-1+V(1))$ oscillating in the region $x \in (0,1)$, which are not present for $V(1)<2$.
Rather surprisingly, the transition at $V(1)=2$ is not accompanied or highlighted by any sudden local change of the Hamiltonian flow, albeit a global change of the topology is present. This can be understood also by noticing that the ordered set of values of the critical energies as a function of $V(1)$ features a crossing for $V(1) = 2$.
As a consequence, there is a critical trajectory connecting the two hyperbolic points at $x=0$ and $x=1$ respectively. We show this mechanism in Fig.~\ref{fig:phase_space_generic}.

The presence of Stark localized orbits can be actually detected by making use of a topological invariant. In fact, consider in phase space a periodic trajectory $(x(t),k(t))$ with period $T$ (which, in general, depends on the trajectory) and define the winding number
\be\label{eq:top_inv}
\mathscr{N} = \oint\frac{dk}{2\pi} \equiv \frac{1}{2\pi}\int_0^T dt \  \dot{k}(t).
\ee
This quantity corresponds to the number of times a trajectory winds around the Brillouin zone, as we explain below.
We first observe that the integral defining $\mathscr{N}$ in Eq.~\eqref{eq:top_inv} is invariant under a time reparameterization $t \mapsto \tau = \tau(t)$ of the trajectory, and thus it depends only on the shape of the trajectory.
Moreover, the invariance of the integral under small deformation of the trajectory
$k(t) \to k(t) + \delta k (t)$ follows from the fact that $k$ behaves like an angle variable and $\mathscr{N}$ is a winding number. Stated more formally,
$dk$ is a closed 1-form (as $d^2k=0$), but it can be different from zero as $dk$ is not an exact differential.
Indeed, strictly speaking, $k$ is not a smooth function of time, being defined up to $2\pi{\mathbb Z}$.
As an example, let us calculate $\mathscr{N}$  in Eq.~\eqref{eq:top_inv} on a closed trajectory with turning points $x_0$ and $x_1$.
We parametrize the integral with the spatial variable $x$ and, denoting by $E$ the (conserved) energy of the trajectory and by $T(E)$ the corresponding period, we get
\be
\begin{split}
\mathscr{N} &= \frac{1}{2\pi}\int_0^{T(E)}\! dt \,  \dot{k}(t) \\
&= -\frac{1}{\pi}\int_{x_0}^{x_1} \!dx\,\frac{V'(x)}{\sqrt{1-[V(x)-E]^2}} = \frac{1}{\pi}\left[ \text{arcsin}(V(x_1)-E)-\text{arcsin}(V(x_0)-E)\right],
\end{split}
\ee
where we used Eqs.~\eqref{eq:Ham_V}, \eqref{eq:eq-motion} and the fact that $E = {\mathcal H}(x(t),k(t))$.
Since the velocity vanishes at the turning points, we have that $V(x_{0,1})-E$ can be either $1$ or $-1$. As a consequence, $\mathscr{N} = \pm 1$ --- corresponding to Stark localization --- or $\mathscr{N} = 0$.

\subsection{The harmonic potential}

\begin{figure}
    \centering
    \includegraphics[width=0.6\textwidth]{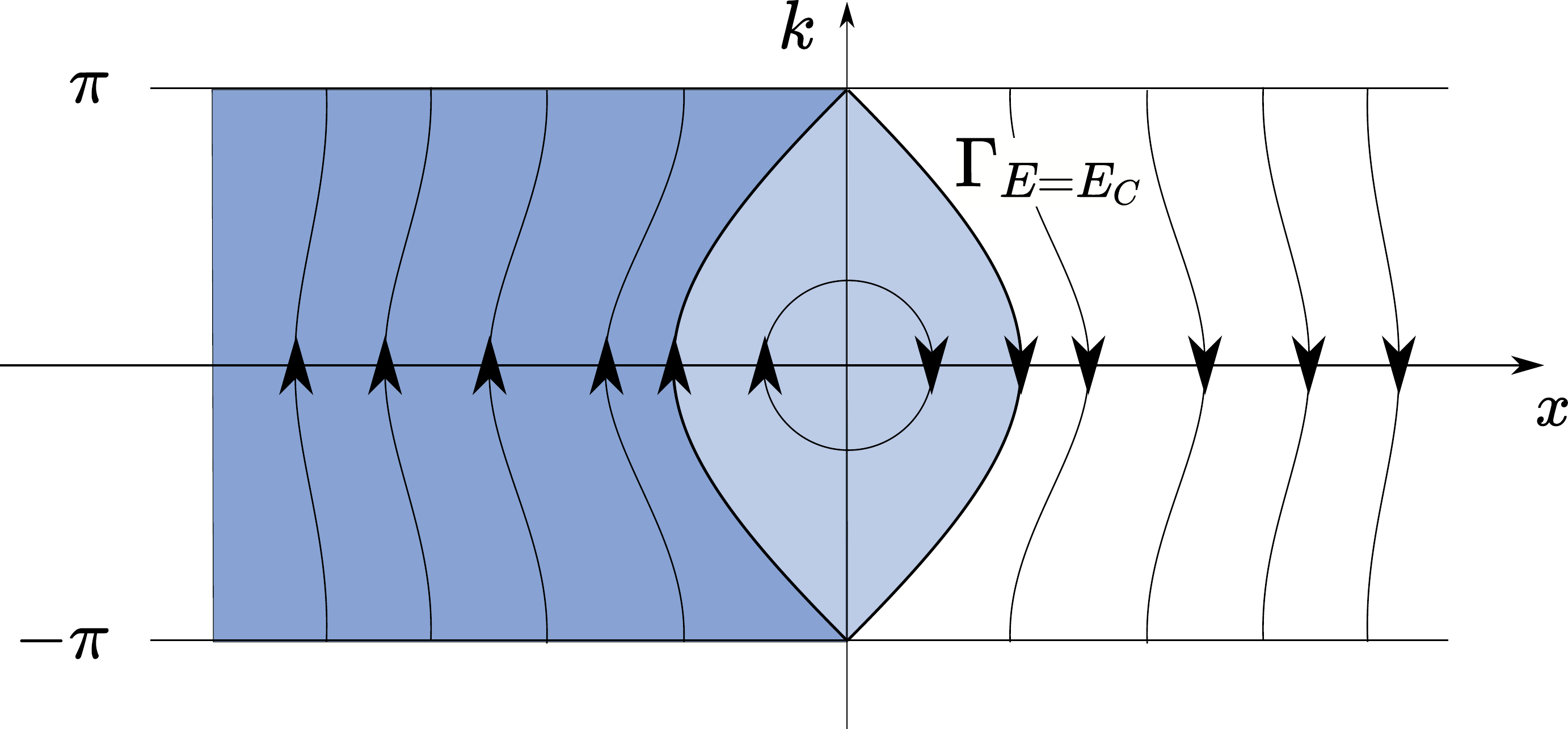}
    \caption{Trajectories in phase space of the Hamiltonian \eqref{eq:evolut_n} with the quadratic potential in Eq.~\eqref{eq:quad-pot}. The critical line $\Gamma_{E=E_c}$ separates the region of phase space with $E<E_c$, in which oscillations occur, from the one with $E>E_c$, where the trajectories Stark localized. 
    For a domain wall initially localized at $x = 0$,
    the long-time average $n_\infty(x,k)$ of the occupation number $n(x,k;t)$ calculated according to Eq.~\eqref{eq:wign_func_quadr} equals 1 for $(x,k)$ belonging to the region colored in darker azure, 1/2 within the region colored in lighter azure, and 0 otherwise.
    }
    \label{fig:Wig_funct_quadr}
\end{figure}

We finally discuss in detail the case of the harmonic potential~\cite{csc-13,csc-13/2}, relating the semi-classical GHD predictions with the microscopic model in Eq.~\eqref{eq:hamiltonian}. In particular, 
we consider
\be
\label{eq:quad-pot}
V(x) = \frac{1}{2}\l \frac{x}{\xi}\r^2,
\ee
where $\xi$ plays the role of a typical length, assumed to be much larger than the lattice spacing (i.e., $\xi \gg 1$).
The critical points of the Hamiltonian ${\cal H}(x,k)$ with this potential are $(x_c,k_c) = (0,0)$, which is elliptic and corresponds to the minimal energy $-1$, and $(x_c,k_c) = (0,\pi)$, which is hyperbolic and corresponds to $E_c=1$. 
Accordingly, for $E>E_c$ one observes Stark localization of the trajectories, while the usual oscillations --- which would  be present also in the absence of the lattice, i.e., with $-\cos k$ in Eq.~\eqref{eq:Ham_V} replaced by $-1+k^2/2$ --- arise for $-1<E<E_c$. 
For later convenience, we write down explicitly the set of points belonging to the critical isoenergetic line (see Eq.~\eqref{eq:Energy_Set}), i.e., 
\be
\Gamma_{E=E_c} = \{ (x,k) \,  | \, x=\pm 2\xi \cos(k/2) \quad\mbox{with}\quad k\in [-\pi,\pi]\}.
\ee
The various type of trajectories in phase space are illustrated in Fig.~\ref{fig:Wig_funct_quadr}. 
While for generic potentials it is not possible, in general, to proceed further, in this case we can actually make quantitative predictions for the time evolution.
For this purpose, we focus on the dynamics of the domain-wall state \eqref{eq:dwall_state} and we study the local occupation number $n(x,k;t)$ and the value it takes along the classical trajectories.
First, we observe that, for $E>E_c=1$, the surface $\Gamma_E$ in Eq.~\eqref{eq:Energy_Set} contains two disjoint trajectories, which wind around the Brillouin zone and belong to the half-plane $x>0$ and $x<0$, respectively (see Fig.~\ref{fig:Wig_funct_quadr}). Since, at the initial time, these trajectories are either completely empty for $x>0$ (i.e., $n(x<0,k;t=0)=1$) or filled for $x>0$ (i.e., $n(x>0,k;t=0)=0$), the corresponding dynamics is simply given by 
\be
n(x,k;t) = n(x,k;t=0), \quad \mbox{for} \quad (x,k) \in \Gamma_{E>E_c}.
\label{eq:n-inf-grt}
\ee
We now consider $E<E_c$ (with $E>-1$), for which $\Gamma_{E}$ contains a single trajectory, initially filled for $x<0$ and empty for $x>0$. 
While an  exact description of the dynamics at all times $t>0$ is possible, we focus here on the long-time average $n_\infty$ of the occupation number $n$, defined as
\be
\label{eq:wign_func_quadr}
n_\infty(x,k) \equiv \underset{t\rightarrow \infty}{\lim}\frac{1}{t}\int^{t}_0 \!dt'\, n(x,k;t').
\ee
In this way, the occupation along a trajectory, after this averaging, takes its mean value and therefore 
\be
n_\infty(x,k) = 1/2, \quad \mbox{for} \quad (x,k) \in \Gamma_{E<E_c},
\label{eq:n-inf-less}
\ee
independently of the actual value of $E$. 
The resulting value of $n_\infty(x,k)$ in phase space in indicated in Fig.~\ref{fig:Wig_funct_quadr}: the darker azure region corresponds to $n_\infty=1$, the lighter azure region to $n_\infty=1/2$, and the white region to $n_\infty=0$.

The time-averaged spatial density $\rho_\infty(x)$ for a certain value of $x$ can then be obtained by integrating this $n_\infty(x,k)$ over the momentum $k$. 
This yields
\be
\label{eq:rho_inf_quadr}
\rho_\infty(x) \equiv \underset{t\rightarrow \infty}{\lim}\frac{1}{t}\int^{t}_0 \!dt'\, \rho(x,t') = \int_{-\pi}^{\pi}\frac{dk}{2\pi}n_\infty(x,k) = \begin{cases} 1 \quad\mbox{for}\quad x<-2\xi, \\ 
\frac{1}{\pi}\text{arccos}\l \frac{x}{2\xi}\r \quad\mbox{for}\quad |x| < 2 \xi,\\
0 \quad\mbox{for}\quad x> 2\xi.
\end{cases}
\ee

\begin{figure}[t]
    \centering
    \includegraphics[width=0.7\textwidth]{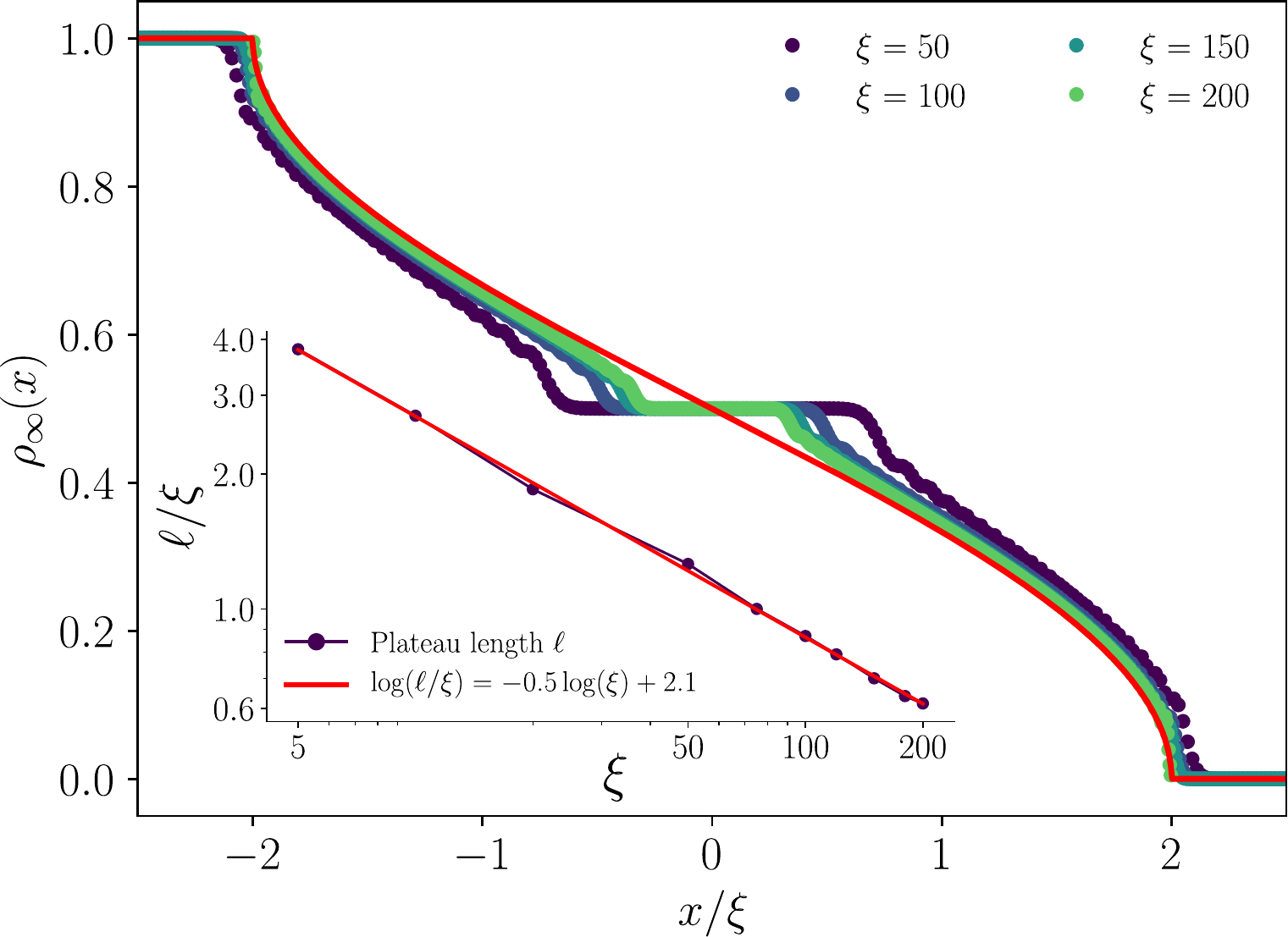}
    \caption{Infinite-time averaged density $\rho_\infty(x)$ [see Eq.~\eqref{eq:rho_inf_quadr}] as a function of $x/\xi$ for the quadratic potential in Eq.~\eqref{eq:quad-pot} and various values of $\xi$, starting from a domain wall localized at $x=0$.
    Symbols correspond to the numerical data obtained by computing the time average using the diagonal ensemble (see Appendix~\ref{sec:app:numerics}) 
    on a chain of length $L=800$. The red solid curve indicates  the analytical prediction in Eq.~\eqref{eq:rho_inf_quadr}, which the numerical data approach as $\xi$ increases. In particular, the plateau of extension $\ell$ which is clearly visible in the numerical data for $x$ around 0 vanishes. In fact, the inset shows the decays of $\ell/\xi$ as $\xi$ increases, with a good fit  $\ell/\xi \sim \xi^{-1/2}$. This indicates that, on the scale of $\xi$, the plateau vanishes as $\xi$ increases.}
    \label{fig:rho_quadr}
\end{figure}

Figure~\ref{fig:rho_quadr} shows the time-averaged density profile $\rho_\infty(x)$  as a function of $x/\xi$, as obtained numerically (symbols) for various values of $\xi$ [see Appendix~\ref{sec:app:numerics} and, in particular Eq.~\eqref{app:eq:inf_time_limit} therein]. These data show data collapse upon increasing $\xi$ and the resulting master curve agrees with the one predicted on the basis of GHD in Eq.~\eqref{eq:rho_inf_quadr}, reported as a solid line. Interestingly, the numerical data are characterized by the presence of a plateau around $x=0$, which is not predicted by GHD. However, the spatial extension $\ell$ of this plateau turns out to grow slower than $\xi$ upon increasing it, with $\ell/\xi \sim \xi^{-1/2}$, as shown in the inset of the figure. Accordingly, in the limit $\xi \rightarrow \infty$ we are interested in, with $x/\xi$ kept fixed, the plateau effectively vanishes and the prediction in Eq.~\eqref{eq:rho_inf_quadr} is recovered.
The presence of this plateau can be actually explained as follows. At a given energy $E$ slightly above $E_c=1$, there are two Stark-localized trajectories (one for $x<0$ and the other for $x>0$) which, according to the semi-classical equation of motion, are dynamically disconnected.
However, when these trajectories approach each other in the vicinity of the hyperbolic critical point $(x_c,k_0) = (0,\pi)$, quantum effects may mix them via quantum tunneling. This tunneling is expected to be suppressed as the two trajectories further separate in space and therefore it should occur predominantly for $E\gtrsim E_c$ and $x\simeq 0$. Because of this tunneling, the resulting value of $n_\infty$ would be the average $1/2$ of the values that $n_\infty$ would have on the two separate branches, in contrast to the semi-classical prediction and in agreement with the presence of the plateau in Fig.~\ref{fig:rho_quadr}. Beyond this heuristic explanation, however, a quantitative study of this tunneling is beyond the scopes of the present work.

\section{Conclusions and outlook}
\label{sec:Conclusion}

In this work, we investigated the dynamics of the  Fermi gas on a lattice in the presence of an external potential, and we study the phenomenon of Stark localization by using the approaches provided by generalized hydrodynamics (GHD) and quantum GHD. 
In particular, considering the case of a linear potential, we derive analytical predictions for the evolution of an initial domain-wall state. 
We compare these predictions with exact numerical computations at finite number of particles, finding perfect agreement in the thermodynamic limit. 
In the presence of a generic potential, we analyze the mechanism which is responsible for the localization of the particles. This analysis shows that the occurrence of localization does not require a fine-tuned external potential but it rather hinges on having a bounded kinetic energy, characterized by a finite band.
Moreover, we argue that the topology of the classical trajectories in the phase space of the system is the key feature which determines the possible presence of Stark localization.
As an illustrative example, we consider the dynamics in the presence of a quadratic potential --- usually not discussed in the context of Stark localization --- and show the agreement between our description and the results of numerical calculations.

We expect that GHD, which allowed us to derive easily the predictions presented in this work, should be able to describe accurately the dynamics and the possible occurrence of Stark localization in generic settings. In particular, modifications of the kinetic term, as long as they span a finite band, can be investigated as described in this work and they are expected to result in a similar phenomenology. 
However, the most important generalization of the approach discussed here would be towards the study of integrable interacting models on the lattice, such as the XXZ model~\cite{scd-22,bcdnf-16,cdlcd-20}. GHD methods are powerful enough to provide analytical predictions also in this case, although the calculations are significantly more challenging than those reported here because of the increasing complexity of the corresponding hydrodynamic equations.
However, it is reasonable to expect that, with some effort, exact predictions can be obtained also for the interacting case, shedding some light on the phenomenon of many-body Stark localization \cite{shmp-19,nbr-19,dgp-21}. 

We emphasize that our analysis requires that the potential varies on  spatial scales which are large compared to the lattice spacing, so that the condition of applicability of the GHD is met. However, one might wonder whether it is possible to relax this assumption in order to describe, e.g., localized potentials arising from defects or impurities. While some specific protocols have been considered and some progress has been made in this direction~\cite{csrc-23,lsp-19,rc-23}, a general theory is still lacking. In the case considered in this work, a significant difficulty that hinders a straightforward application of the GHD approach, is the presence of long-range correlation generated at these defects by  the quantum scattering and the subsequent ballistic spread across the system. We plan to address this problem in the future.

\section*{Acknowledgement}
The authors are grateful to Federico Balducci and Stefano Scopa for a careful reading of the manuscript. LC and CV are also grateful to Federico Rottoli and Stefano Scopa for useful discussions. CV and AG would like to thank Federico Balducci, Alessio Lerose, and Antonello Scardicchio for collaboration on related projects. PC and LC acknowledge support by the ERC under Consolidator grant number 771536 NEMO. AG acknowledges financial support from the PNRR MUR project PE0000023-NQSTI. CV thanks ICTP for hospitality.

\begin{appendix}

\section{Derivation of the semi-classical Hamiltonian}
\label{app:sec:hamilt}

In this Appendix we recall how to derive the semi-classical Hamiltonian in Eq.~\eqref{eq:Class_Ham} starting from that of the original quantum chain in Eq.~\eqref{eq:hamiltonian}.
The fundamental step consists in passing from a second-quantized to a first-quantized form of the operators appearing in the Hamiltonian. In this respect, consider an operator $\hat{O}$ which can be written as $\hat{O} = \sum_{i=1}^N \hat{o}(i)$, where each operator $\hat{o}(i)$ acts on the one-particle subspace of the $i$-th particle (with $i\in\{1,\ldots,N\}$). Then in second quantization, i.e., in Fock space, $\hat{O}$ is written as $\hat{O} = \sum_{r,s} c_r^{\dagger} \bra{r} \hat{o} \ket{s} c_s$, where $c_r^{\dagger}$ and $c_s$ are, respectively, the creation and annihilation operators for a particle in the state $\ket{r}$ and $\ket{s}$, being $\ket{r}$ and $\ket{s}$ elements of a generic orthonormal basis of the Hilbert space.
In order to derive the semiclassical Hamiltonian, we have first to perform the opposite change of basis, starting from the knowledge of the matrix elements $\bra{r} \hat{o} \ket{s}$. In turn, the latter can be conveniently derived by diagonalizing the Hamiltonian in Eq.~\eqref{eq:hamiltonian}, which is quadratic. Focussing, first, on the kinetic term, it is convenient to introduce the operators $c_p$ and $c^\dagger_p$ in momentum space as
\be
    \psi_x = \frac{1}{\sqrt{L}} \sum_p e^{ipx} c_p \quad \mbox{and} \quad  \psi_x^{\dagger} = \frac{1}{\sqrt{L}} \sum_p e^{-ipx} c_p^{\dagger}. 
\ee
where $p= 2\pi n/L$, $n\in \mathbb{Z}$.
A simple substitution leads to
\be
    \frac{1}{2} \sum_x (\psi_x^{\dagger} \psi_{x+1} + h.c.) = \sum_{p} (\cos{p}) \, c_{p}^{\dagger} c_p,
\ee
which can be equivalently written~\cite{n-98-book} as $\sum_{i=1}^N \cos \hat{p}_i$, where $\hat{p}_i$ the momentum operator defined in the one-particle subspace of the $i$-th particle.
The second term in the Hamiltonian in Eq.~\eqref{eq:hamiltonian} is already in diagonal form, and therefore we have
\be
    \sum_x V(x) \psi_x^{\dagger} \psi_x = \sum_{i=1}^N V(\hat{x}_i),
\ee
where $\hat{x}_i$ the position operator defined in the one-particle subspace of the $i$-th particle.
Accordingly, the Hamiltonian can be written in terms of the fundamental one-particle operators $\{\hat x_i, \hat p_i \}_i$ (i.e., in the form of the first quantization) as
\be
\label{app:eq:ham_first_quant}
    H =  \sum_{i=1}^N \left[ - \cos{\hat{p}_i} + V(\hat{x}_i) \right].
\ee
The semiclassical approximation of Eq.~\eqref{app:eq:ham_first_quant} can now be done as usual, by substituting the quantum operators with the corresponding classical variables in phase space, leading to the Hamiltonian in Eq.~\eqref{eq:Ham_V}.

\section{Numerics}
\label{sec:app:numerics}

In this appendix, following Refs.~\cite{p-03,p-04,p-12,pe-09,cp-01,pkl-99}, we briefly explain the numerical methods employed in order to study the dynamics of the Hamiltonian in Eq.~\eqref{eq:hamiltonian}. The correlation matrix $C(0)$ of a Gaussian state, defined in Eq.~\eqref{eq:Cmatrix}, evolves as
\be
C(t) = e^{i\hat{H}t}C(0)e^{-i\hat{H}t},
\ee
where $\hat{H}$ is the single-particle Hamiltonian defined by
\be
H = \sum_{x,x'} \hat{H}_{x,x'} \psi^\dagger_x \psi_{x'}.
\ee
Given $C(t)$, the particle density $\rho(x,t)$ and the particle current $j(x,t)$ are easily recovered from the definitions in  Eqs.~\eqref{eq:density} and \eqref{eq:current}.
In order to compute the R\'enyi entropies of a sublattice $A$, we first need to project $C(t)$ over $A$, obtaining the restricted matrix
\be
\l C_A(t)\r_{x,x'} \equiv \l C(t)\r_{x,x'}, \quad x,x' \in A.
\ee
Then, one can show~\cite{ac-17,p-12} that the $n$-th R\'enyi entropy is expressed by
\be
S_n(t) = \frac{1}{1-n}\text{Tr}\log \l C^n_A(t) + (1-C_A(t))^n\r.
\ee
The long-time average $C_\infty$ of $C(t)$, defined as
\be
C_\infty \equiv \lim_{T \to \infty} \frac{1}{T} \int_0^T dt \, C(t),
\ee
can be easily computed, once the eigenvalues of $\hat{H}$ are known. In fact, assuming that the spectrum of $\hat{H}$ is non-degenerate, a straightforward algebra yields
\be
C_\infty = \sum_E \ket{E}\bra{E}C(0)\ket{E}\bra{E},
\ee
where $\{\ket{E}\}_E$ is the set of eigenvectors of $\hat{H}$.
In other words,  the correlation function at long times thermalizes in average to its diagonal ensemble, as the oscillations around that value --- due to transitions among distinct eigenvectors --- are averaged out.
From $C_\infty$ one can extract directly the long-time average of one-particle observables. For example, the average density of particles $\rho_{\infty}(x) $ can be expressed in terms of $C_\infty$ as 
\be
\label{app:eq:inf_time_limit}
\rho_{\infty}(x) \equiv \l C_\infty\r_{x,x}.
\ee

\section{Exact results on the lattice}
\label{sec:app:lattice}

In this Appendix we briefly discuss how the GHD predictions discussed in Sec.~\ref{sec:GHD_Stark} for the lattice Hamiltonian \eqref{eq:hamiltonian} with the linear potential $V(x) = -hx$ can be recovered from its exact evolution discussed, e.g., in Ref.~\cite{bglsv-23}. 
For clarity, we will denote by capital letters the positions on the lattice, which therefore assume only integer values. The corresponding variables on the continuum, instead, will be denoted by lowercase letters. 
Let us start from the density $\rho(x,t)$: 
for the domain-wall initial state we are considering in this work [see Eq.~\eqref{eq:dwall_state}], the exact expression on an infinite chain is \cite{bglsv-23}
\be
\label{eq:rho_lattice1}
\rho (X,t) = \sum_{Y \geq X} J_Y^2 \left( l(t) \right) = \int dX' \, n(X') \, J_{|X-X'|}^2 \left( l(t) \right).
\ee
Here $l(t)$ is given by Eq.~\eqref{eq:loc-length}, $J_Y(l)$ is the Bessel function of the first kind, while $n(X') = \sum_{X''=-\infty}^{0} \delta(X'-X'')$ has been introduced for later convenience and corresponds to the distribution of particles in the initial state. The expression in Eq.~\eqref{eq:rho_lattice1} can be obtained straightforwardly by diagonalizing the Hamiltonian and by computing the expectation value $\rho(x,t) = \bra{\Psi_0} \psi_x^{\dagger} \psi_x \ket{\Psi_0}$ using, e.g., Wick theorem (see Ref.~\cite{bglsv-23} for further details). 
The integral representation introduced in Eq.~\eqref{eq:rho_lattice1} in terms of $n(X')$ is useful for taking the continuum limit of vanishing lattice sapcing $a$, which is suitably obtained by considering the continuous variable $x = aX$ and by rescaling $t \to t/a$ (as to access long times in the formal limit $a$) and $h \to ah$ (weak field), so that $l(t) \to l(t)/a$. 
With these substitutions and the change of variables $X' = x'/a$, one gets
\be
    n\left( X' = x'/a \right) = a \sum_{X''=-\infty}^{0} \delta(x'-aX'') \stackrel{a\to 0}{\longrightarrow} \int_{-\infty}^{0} \, dx''\, \delta(x'-x'') = \theta(-x'),
\ee
where $\theta(x)$ is the unit step function.
Inserting this expression into Eq.~\eqref{eq:rho_lattice1} and using the asymptotic expansion of Bessel functions $J^2_x(y) \simeq \theta (y-x)\frac{1}{\pi} \frac{1}{y\sqrt{1-(x/y)^2}}$~\cite{NIST:DLMF} for $x$, $y \to \infty$, with fixed $x/y$, one gets \cite{bglsv-23} 
\begin{align}
    \rho(x,t) \stackrel{a\to 0}{\longrightarrow}  \int_{-\infty}^{+\infty} dx' \, \theta(-x')\, \frac{\theta(l(t)-|x-x'|)}{\pi l(t) \sqrt{1-|x-x'|^2/l(t)^2}} =  \int_{-1}^{1} dy\,\frac{\theta(-y \, l(t) - x)}{\pi\sqrt{1-y^2}}&\nonumber\\
    =\int_{-1}^{-x/l(t)} dy\,\frac{1}{\pi\sqrt{1-y^2}} = \frac{1}{\pi} \arccos{\left( \frac{x}{l(t)} \right)}&,
\end{align}
(in the last line we assume that $|x| \le l(t)$, otherwise the integral vanishes if $x\geq l(t)$ or equals 1 if $x\leq -l(t)$) which indeed renders the expression in Eq.~\eqref{eq:exp-j}, obtained via GHD. 

A similar procedure can be followed also for the particle current on the lattice, which we denote by $j(x,t)_{\mathrm{lattice}}$ and which is defined as in Eq.~\eqref{eq:current}; its expression on the lattice is given by~\cite{bglsv-23}
\be
    \label{eq:current_lattice_result}
    j(x,t)_{\mathrm{lattice}} = \frac{\sin(ht)}{2h} \left[ J_x^2\left( \frac{2}{h} \sin{\frac{ht}{2}} \right) - J_{x+1}\left( \frac{2}{h} \sin{\frac{ht}{2}} \right) J_{x-1}\left( \frac{2}{h} \sin{\frac{ht}{2}} \right) \right].
\ee
For simplicity, we consider $x=0$ (being the case $x\neq 0$ completely analogous) and we introduce the lattice spacing $a$ as described above, obtaining
\be
    \label{eq:current_lattice_limit}
    j(0,t)_{\mathrm{lattice}} = \frac{\sin(ht)}{2ah} \left[ J_0^2\left( \frac{2}{ah} \sin{\frac{ht}{2}} \right) + J_{1}^2\left( \frac{2}{ah} \sin{\frac{ht}{2}} \right) \right],
\ee
where we used the property $J_{-n}(x) = (-1)^n J_n(x)$ of the Bessel function with $n \in \mathbb{Z}$.
Looking at the arguments of the Bessel functions, it might seem that the current  has a period $T$ such that  $hT/2=2\pi$. However, using the previous property and the fact that $J_n(-x) = J_{-n}(x)$ for $n \in \mathbb{Z}$, one easily proves that $J_0^2(x) + J_1^2(x) = J_0^2(-x) + J_1^2(-x)$ and therefore  the actual period of $j(0,t)_{\mathrm{lattice}}$ is such that $hT/2 = \pi$, i.e., $T$ is given by Eq.~\eqref{eq:expT}. 
Accordingly, we can focus the attention on a single period and by using the asymptotic expansion of Bessel functions $J_{x}(\gamma)$ for large argument $\gamma \to +\infty$ at fixed order $x$, i.e.,  $J_{x}(\gamma) \simeq \sqrt{\frac{2}{\pi \gamma}} \cos{\left( \gamma -\frac{\pi}{2} x - \frac{\pi}{4} \right)}$ \cite{NIST:DLMF}, one gets (assuming $h>0$)
\begin{align}
    \label{eq:current_lattice_cont}
    j(0,t)_{\mathrm{cont.}} &= \frac{\sin(ht)}{2ah}  \frac{ah}{\pi |\sin(ht/2)|} \left[ \cos^2\left( \frac{2}{ah} \sin{\frac{ht}{2}} -\frac{\pi}{4} \right) + \cos^2\left( \frac{2}{ah} \sin{\frac{ht}{2}} -\frac{3\pi}{4} \right) \right]\\
    &= \frac{\sin(ht)}{2\pi |\sin(ht/2)|} =  \frac{1}{\pi} \cos(ht/2)\; {\rm sgn}(\sin(ht/2)),
\end{align}
where ${\rm sgn}(x) \equiv x/|x|$ is the sign function. This expression coincides with the prediction of GHD reported in Eq.~\eqref{eq:exp-j}, specialized to the case $x=0$. 
Note that the discontinuity of the previous expression at times $t_k$ such that $ht_k/2 = k \pi$ with $k\in \mathbb{Z}$ (which correspond to the change of sign of $\sin(ht/2)$), already emphasized after Eq.~\eqref{eq:eqrho0}, emerges only in the limit $a\to 0$, while the function turns out to be regular for finite values of $a$, see also the left panel of Fig.~\ref{fig:current}.

\end{appendix}

\printbibliography

@article{bcdnf-16,
  title = {Transport in Out-of-Equilibrium {XXZ} Chains: Exact Profiles of Charges and Currents},
  author = {Bertini, Bruno and Collura, Mario and De Nardis, Jacopo and Fagotti, Maurizio},
  journal = {Phys. Rev. Lett.},
  volume = {117},
  issue = {20},
  pages = {207201},
  numpages = {8},
  year = {2016},
  month = {11},
  doi = {10.1103/PhysRevLett.117.207201},
}

@article{wck-13,
  title = {Hydrodynamic description of hard-core bosons on a {G}alileo ramp},
  author = {Wendenbaum, Pierre and Collura, Mario and Karevski, Dragi},
  journal = {Phys. Rev. A},
  volume = {87},
  issue = {2},
  pages = {023624},
  numpages = {7},
  year = {2013},
  month = {2},
  publisher = {American Physical Society},
  doi = {10.1103/PhysRevA.87.023624}
}

@article{csc-13,
  title = {Equilibration of a {T}onks-{G}irardeau Gas Following a Trap Release},
  author = {Collura, Mario and Sotiriadis, Spyros and Calabrese, Pasquale},
  journal = {Phys. Rev. Lett.},
  volume = {110},
  issue = {24},
  pages = {245301},
  numpages = {5},
  year = {2013},
  month = {6},
  publisher = {American Physical Society},
  doi = {10.1103/PhysRevLett.110.245301}
}

@article{csc-13/2,
doi = {10.1088/1742-5468/2013/09/P09025},
year = {2013},
month = {9},
publisher = {IOP Publishing and SISSA},
volume = {2013},
number = {09},
pages = {P09025},
author = {Mario Collura and Spyros Sotiriadis and Pasquale Calabrese},
title = {Quench dynamics of a {T}onks–{G}irardeau gas released from a harmonic trap},
journal = {J. Stat. Mech.: Theor. Exp.}
}

@article{kmm-18,
  title = {Quantum quench and thermalization of one-dimensional {F}ermi gas via phase-space hydrodynamics},
  author = {Kulkarni, Manas and Mandal, Gautam and Morita, Takeshi},
  journal = {Phys. Rev. A},
  volume = {98},
  issue = {4},
  pages = {043610},
  numpages = {15},
  year = {2018},
  month = {10},
  publisher = {American Physical Society},
  doi = {10.1103/PhysRevA.98.043610}
}

@article{cadt-16,
  title = {Emergent Hydrodynamics in Integrable Quantum Systems Out of Equilibrium},
  author = {Castro-Alvaredo, Olalla A. and Doyon, Benjamin and Yoshimura, Takato},
  journal = {Phys. Rev. X},
  volume = {6},
  issue = {4},
  pages = {041065},
  numpages = {17},
  year = {2016},
  month = {12},
  doi = {10.1103/PhysRevX.6.041065},
}

@article{rcdd-20,
  title = {Quantum Generalized Hydrodynamics},
  author = {Ruggiero, Paola and Calabrese, Pasquale and Doyon, Benjamin and Dubail, J\'er\^ome},
  journal = {Phys. Rev. Lett.},
  volume = {124},
  issue = {14},
  pages = {140603},
  numpages = {7},
  year = {2020},
  month = {4},
  publisher = {American Physical Society},
  doi = {10.1103/PhysRevLett.124.140603},
}

@article{ccad-07,
	doi = {10.1007/s10955-007-9422-x},
	year = {2007},
	month = {10},
	publisher = {Springer Science and Business Media {LLC}},  
	volume = {130},  
	number = {1},  
	pages = {129--168},
	author = {J. L. Cardy and O. A. Castro-Alvaredo and B. Doyon},
	title = {Form Factors of Branch-Point Twist Fields in Quantum Integrable Models and Entanglement Entropy},
	journal = {J. Stat. Phys.}
}

@article{rcdd-21,
  doi = {10.1088/1751-8121/ac3d68},
  title={Quantum generalized hydrodynamics of the {T}onks-{G}irardeau gas: density fluctuations and entanglement entropy},
  author={Ruggiero, Paola and Calabrese, Pasquale and Doyon, Benjamin and Dubail, J{\'e}r{\^o}me},
  journal={J. Phys. A: Math. Theor.},
  volume={55},
  number={2},
  pages={024003},
  year={2021},
  publisher={IOP Publishing}
}

@article{sc-14,
doi = {10.1088/1742-5468/2014/07/P07024},
year = {2014},
month = {7},
publisher = {IOP Publishing and SISSA},
volume = {2014},
number = {7},
pages = {P07024},
author = {Spyros Sotiriadis and Pasquale Calabrese},
title = {Validity of the GGE for quantum quenches from interacting to noninteracting models},
journal = {J. Stat. Mech.: Theor. Exp.},
}

@article{rdyo-07,
  title = {Relaxation in a Completely Integrable Many-Body Quantum System: An Ab Initio Study of the Dynamics of the Highly Excited States of 1D Lattice Hard-Core Bosons},
  author = {Rigol, Marcos and Dunjko, Vanja and Yurovsky, Vladimir and Olshanii, Maxim},
  journal = {Phys. Rev. Lett.},
  volume = {98},
  issue = {5},
  pages = {050405},
  numpages = {4},
  year = {2007},
  month = {2},
  publisher = {American Physical Society},
  doi = {10.1103/PhysRevLett.98.050405},
}

@article{pssv-11,
  title = {Colloquium: Nonequilibrium dynamics of closed interacting quantum systems},
  author = {Polkovnikov, Anatoli and Sengupta, Krishnendu and Silva, Alessandro and Vengalattore, Mukund},
  journal = {Rev. Mod. Phys.},
  volume = {83},
  issue = {3},
  pages = {863--883},
  numpages = {0},
  year = {2011},
  month = {8},
  publisher = {American Physical Society},
  doi = {10.1103/RevModPhys.83.863}
}

@article{bdl-18,
  title = {Nonequilibrium Steady State Generated by a Moving Defect: The Supersonic Threshold},
  author = {Bastianello, Alvise and De Luca, Andrea},
  journal = {Phys. Rev. Lett.},
  volume = {120},
  issue = {6},
  pages = {060602},
  numpages = {7},
  year = {2018},
  month = {2},
  publisher = {American Physical Society},
  doi = {10.1103/PhysRevLett.120.060602}
}

@article{skcd-21,
doi = {10.1088/1751-8121/ac20ee},
year = {2021},
month = {9},
publisher = {IOP Publishing},
volume = {54},
number = {40},
pages = {404002},
author = {Stefano Scopa and Alexandre Krajenbrink and Pasquale Calabrese and Jérôme Dubail},
title = {Exact entanglement growth of a one-dimensional hard-core quantum gas during a free expansion},
journal = {J. Phys. A: Math. Theor.},
}

@article{cdlcd-20,
  title = {Domain wall melting in the spin-$\frac{1}{2}$ {XXZ} spin chain: Emergent {L}uttinger liquid with a fractal quasiparticle charge},
  author = {Collura, Mario and De Luca, Andrea and Calabrese, Pasquale and Dubail, J\'er\^ome},
  journal = {Phys. Rev. B},
  volume = {102},
  issue = {18},
  pages = {180409},
  numpages = {8},
  year = {2020},
  month = {11},
  publisher = {American Physical Society},
  doi = {10.1103/PhysRevB.102.180409},
}

@article{abfpr-21,
doi = {10.1088/1742-5468/ac257d},
year = {2021},
month = {11},
publisher = {IOP Publishing and SISSA},
volume = {2021},
number = {11},
pages = {114004},
author = {Vincenzo Alba and Bruno Bertini and Maurizio Fagotti and Lorenzo Piroli and Paola Ruggiero},
title = {Generalized-hydrodynamic approach to inhomogeneous quenches: correlations, entanglement and quantum effects},
journal = {J. Stat. Mech.: Theor. Exp.},
}

@article{d-20,
	title={{Lecture notes on Generalised Hydrodynamics}},
	author={Benjamin Doyon},
	journal={SciPost Phys. Lect. Notes},
	pages={18},
	year={2020},
	publisher={SciPost},
	doi={10.21468/SciPostPhysLectNotes.18},
}

@article{e-22,
	title = {A short introduction to Generalized Hydrodynamics},
	issn = {0378-4371},
	doi = {https://doi.org/10.1016/j.physa.2022.127572},
	pages = {127572},
	journal = {Physica A: Stat. Mech. Appl.},
	author = {Essler, Fabian H. L.},
	date = {2022},
}

@article{bd-22,
doi = {10.1088/1742-5468/ac3659},
year = {2022},
month = {1},
publisher = {IOP Publishing and SISSA},
volume = {2022},
number = {1},
pages = {014003},
author = {Isabelle Bouchoule and Jérôme Dubail},
title = {Generalized hydrodynamics in the one-dimensional {B}ose gas: theory and experiments},
journal = {J. Stat. Mech.: Theor. Exp.}
}

@article{dt-17,
  title={A note on generalized hydrodynamics: inhomogeneous fields and other concepts},
  author={Doyon, Benjamin and Yoshimura, Takato},
  journal={SciPost Phys.},
  volume={2},
  number={2},
  pages={014},
  year={2017},
  doi={10.21468/SciPostPhys.2.2.014}
}

@article{dndmp-22,
doi = {10.1088/1742-5468/ac3658},
year = {2022},
month = {1},
publisher = {IOP Publishing and SISSA},
volume = {2022},
number = {1},
pages = {014002},
author = {Jacopo De Nardis and Benjamin Doyon and Marko Medenjak and Miłosz Panfil},
title = {Correlation functions and transport coefficients in generalized hydrodynamics},
journal = {J. Stat. Mech.: Theor. Exp.},
}

@article{bgi-21,
doi = {10.1088/1742-5468/ac12c7},
year = {2021},
month = {8},
publisher = {IOP Publishing and SISSA},
volume = {2021},
number = {8},
pages = {084001},
author = {Vir B Bulchandani and Sarang Gopalakrishnan and Enej Ilievski},
title = {Superdiffusion in spin chains},
journal = {J. Stat. Mech.: Theor. Exp.},
}

@article{bdndl-20,
	title = {Generalized hydrodynamics with dephasing noise},
	volume = {102},
	issn = {2469-9950, 2469-9969},
	doi = {10.1103/PhysRevB.102.161110},
	pages = {161110},
	number = {16},
	journal = {Phys. Rev. B},
	author = {Bastianello, Alvise and De Nardis, Jacopo and De Luca, Andrea},
	date = {2020-10-16}
}

@article{bdlv-21,
	title = {Hydrodynamics of weak integrability breaking},
	volume = {2021},
	issn = {1742-5468},
	doi = {10.1088/1742-5468/ac26b2},
	pages = {114003},
	number = {11},
	journal = {J. Stat. Mech.: Theor. Exp.},
	author = {Bastianello, Alvise and De Luca, Andrea and Vasseur, Romain},
	date = {2021-11-01}
}

@article{bdd-20,
	title = {The effect of atom losses on the distribution of rapidities in the one-dimensional {B}ose gas},
	volume = {9},
	issn = {2542-4653},
	doi = {10.21468/SciPostPhys.9.4.044},
	pages = {044},
	number = {4},
	journal = {{SciPost} Phys.},
	author = {Bouchoule, Isabelle and Doyon, Benjamin and Dubail, Jerome},
	date = {2020-10-02}
}

@article{dnbd-18,
	title = {Hydrodynamic Diffusion in Integrable Systems},
	volume = {121},
	issn = {0031-9007, 1079-7114},
	doi = {10.1103/PhysRevLett.121.160603},
	pages = {160603},
	number = {16},
	journal = {Phys. Rev. Lett.},
	author = {De Nardis, Jacopo and Bernard, Denis and Doyon, Benjamin},
	date = {2018-10-17},
}

@article{dnbd-19,
	title = {Diffusion in generalized hydrodynamics and quasiparticle scattering},
	volume = {6},
	issn = {2542-4653},
	doi = {10.21468/SciPostPhys.6.4.049},
	pages = {049},
	number = {4},
	journal = {{SciPost} Phys.},
	author = {De Nardis, Jacopo and Bernard, Denis and Doyon, Benjamin},
	date = {2019-04-25}
}

@article{mdny-20,
	title = {Diffusion from convection},
	volume = {9},
	issn = {2542-4653},
	doi = {10.21468/SciPostPhys.9.5.075},
	pages = {075},
	number = {5},
	journal = {{SciPost} Phys.},
	author = {Medenjak, Marko and De Nardis, Jacopo and Yoshimura, Takato},
	date = {2020-11-19}
}

@article{ddldnd-21,
	title = {Diffusive hydrodynamics of inhomogenous Hamiltonians},
	volume = {54},
	issn = {1751-8113, 1751-8121},
	doi = {10.1088/1751-8121/ac2c57},
	pages = {494001},
	number = {49},
	journal = {J. Phys. A: Math. Theor.},
	author = {Durnin, Joseph and De Luca, Andrea and De Nardis, Jacopo and Doyon, Benjamin},
	date = {2021-12-10}
}

@article{scd-22,
	title = {Exact hydrodynamic solution of a double domain wall melting in the spin-1/2 {XXZ} model},
	volume = {12},
	issn = {2542-4653},
	doi = {10.21468/SciPostPhys.12.6.207},
	pages = {207},
	number = {6},
	journal = {{SciPost} Phys.},
	author = {Scopa, Stefano and Calabrese, Pasquale and Dubail, Jerome},
	date = {2022-06-28}
}

@article{asw-22,
	title = {Entanglement dynamics of a hard-core quantum gas during a {J}oule expansion},
	volume = {55},
	issn = {1751-8113, 1751-8121},
	doi = {10.1088/1751-8121/ac8209},
	pages = {375301},
	number = {37},
	journal = {J. Phys. A: Math. Theor.},
	author = {Ares, Filiberto and Scopa, Stefano and Wald, Sascha},
	date = {2022-09-16}
}

@article{sh-22,
	title = {Exact hydrodynamic description of symmetry-resolved {R}\'enyi entropies after a quantum quench},
	volume = {2022},
	issn = {1742-5468},
	doi = {10.1088/1742-5468/ac85eb},
	pages = {083104},
	number = {8},
    journal = {J. Stat. Mech.: Theor. Exp.},
	author = {Scopa, Stefano and Horváth, Dávid X},
	date = {2022-08-01}
}

@article{f-20,
	title = {Locally quasi-stationary states in noninteracting spin chains},
	volume = {8},
	issn = {2542-4653},
	doi = {10.21468/SciPostPhys.8.3.048},
	pages = {048},
	number = {3},
    journal = {{SciPost} Phys.},
	author = {Fagotti, Maurizio},
	date = {2020-03-27},
}

@article{sbdd-19,
  title = {Generalized Hydrodynamics on an Atom Chip},
  author = {Schemmer, M. and Bouchoule, I. and Doyon, B. and Dubail, J.},
  journal = {Phys. Rev. Lett.},
  volume = {122},
  issue = {9},
  pages = {090601},
  numpages = {7},
  year = {2019},
  month = {3},
  publisher = {American Physical Society},
  doi = {10.1103/PhysRevLett.122.090601}
}

@article{mzldrw-21,
author = {Neel Malvania  and Yicheng Zhang  and Yuan Le  and Jerome Dubail  and Marcos Rigol  and David S. Weiss },
title = {Generalized hydrodynamics in strongly interacting {1D} {B}ose gases},
journal = {Science},
volume = {373},
number = {6559},
pages = {1129-1133},
year = {2021},
doi = {10.1126/science.abf0147},
}

@article{bglsv-22,
  title = {Localization and Melting of Interfaces in the Two-Dimensional Quantum {I}sing Model},
  author = {Balducci, Federico and Gambassi, Andrea and Lerose, Alessio and Scardicchio, Antonello and Vanoni, Carlo},
  journal = {Phys. Rev. Lett.},
  volume = {129},
  issue = {12},
  pages = {120601},
  numpages = {7},
  year = {2022},
  month = {9},
  publisher = {American Physical Society},
  doi = {10.1103/PhysRevLett.129.120601},
}

@article{bglsv-23,
  title = {Interface dynamics in the two-dimensional quantum {I}sing model},
  author = {Balducci, Federico and Gambassi, Andrea and Lerose, Alessio and Scardicchio, Antonello and Vanoni, Carlo},
  journal = {Phys. Rev. B},
  volume = {107},
  issue = {2},
  pages = {024306},
  numpages = {28},
  year = {2023},
  month = {1},
  publisher = {American Physical Society},
  doi = {10.1103/PhysRevB.107.024306},
}

@article{dsvc-17,
	title={{Conformal field theory for inhomogeneous one-dimensional quantum  systems: the example of non-interacting {F}ermi gases}},
	author={Jérôme Dubail and Jean-Marie Stéphan and Jacopo Viti and Pasquale Calabrese},
	journal={SciPost Phys.},
	volume={2},
	pages={002},
	year={2017},
	publisher={SciPost},
	doi={10.21468/SciPostPhys.2.1.002},
}

@article{cc-09,
doi = {10.1088/1751-8113/42/50/504005},
year = {2009},
month = {12},
publisher = {},
volume = {42},
number = {50},
pages = {504005},
author = {Pasquale Calabrese and John Cardy},
title = {Entanglement entropy and conformal field theory},
journal = {J. Phys. A: Math. Theor.},
}

@article{cc-04,
doi = {10.1088/1742-5468/2004/06/P06002},
year = {2004},
month = {6},
publisher = {},
volume = {2004},
number = {06},
pages = {P06002},
author = {Pasquale Calabrese and  John Cardy},
title = {Entanglement entropy and quantum field theory},
journal = {J. Stat. Mech.: Theor. Exp.},
}

@article{ccd-08,
  doi = {10.1007/s10955-007-9422-x},
  title={Form factors of branch-point twist fields in quantum integrable models and entanglement entropy},
  author={Cardy, John L and Castro-Alvaredo, Olalla A and Doyon, Benjamin},
  journal={J. Stat. Phys.},
  volume={130},
  pages={129--168},
  year={2008},
  publisher={Springer}
}

@article{cc-05,
author = {Calabrese, Pasquale and Cardy, John},
year = {2005},
month = {03},
pages = {P04010},
title = {{Evolution of entanglement entropy in one-dimensional systems}},
volume = {2005},
journal = {J. Stat. Mech.: Theor. Exp.},
doi = {10.1088/1742-5468/2005/04/P04010}
}

@article{jk-04,
	title = {Quantum spin chain, {T}oeplitz determinants and the 
 {F}isher-{H}artwig conjecture},
	volume = {116},
	issn = {1572-9613},
	doi = {10.1023/B:JOSS.0000037230.37166.42},
	pages = {79--95},
	number = {1},
	journal = {J. Stat. Phys.},
	author = {Jin, B.-Q. and Korepin, V. E.},
	date = {2004-08-01},
}

@article{bs-18,
	title = {Characteristic length scales from entanglement dynamics in electric-field-driven tight-binding chains},
	volume = {98},
	issn = {2469-9950, 2469-9969},
	doi = {10.1103/PhysRevB.98.045408},
	pages = {045408},
	number = {4},
	journal = {Phys. Rev. B},
	author = {Bhakuni, Devendra Singh and Sharma, Auditya},
	date = {2018-07-10},
}

@article{w-32,
  title = {On the quantum correction for thermodynamic equilibrium},
  author = {Wigner, E.},
  journal = {Phys. Rev.},
  volume = {40},
  issue = {5},
  pages = {749--759},
  numpages = {0},
  year = {1932},
  month = {6},
  publisher = {American Physical Society},
  doi = {10.1103/PhysRev.40.749}
}

@article{cg-69,
  title = {Density operators and quasiprobability distributions},
  author = {Cahill, K. E. and Glauber, R. J.},
  journal = {Phys. Rev.},
  volume = {177},
  issue = {5},
  pages = {1882--1902},
  numpages = {0},
  year = {1969},
  month = {1},
  publisher = {American Physical Society},
  doi = {10.1103/PhysRev.177.1882}
}

@article{csrc-23,
doi = {10.1209/0295-5075/acb50a},
year = {2023},
month = {1},
publisher = {EDP Sciences, IOP Publishing and Società Italiana di Fisica},
volume = {141},
number = {3},
pages = {31002},
author = {Luca Capizzi and Stefano Scopa and Federico Rottoli and Pasquale Calabrese},
title = {Domain wall melting across a defect},
journal = {Europhys. Lett.}
}

@article{lsp-19,
	title={Non-equilibrium quantum transport in presence of a defect: the  non-interacting case},
	author={Marko Ljubotina and Spyros Sotiriadis and Tomaž Prosen},
	journal={SciPost Phys.},
	volume={6},
	pages={004},
	year={2019},
	publisher={SciPost},
	doi={10.21468/SciPostPhys.6.1.004}
}

@article{pkl-99,
author = {Peschel, Ingo and Kaulke, Matthias and Legeza, Örs},
title = {Density-matrix spectra for integrable models},
journal = {Ann. Phys.},
volume = {511},
number = {2},
pages = {153-164},
doi = {https://doi.org/10.1002/andp.19995110203},
year = {1999}
}

@article{cp-01,
  title = {Density-matrix spectra of solvable fermionic systems},
  author = {Chung, Ming-Chiang and Peschel, Ingo},
  journal = {Phys. Rev. B},
  volume = {64},
  issue = {6},
  pages = {064412},
  numpages = {7},
  year = {2001},
  month = {7},
  publisher = {American Physical Society},
  doi = {10.1103/PhysRevB.64.064412}
}

@article{p-03,
doi = {10.1088/0305-4470/36/14/101},
year = {2003},
month = {3},
volume = {36},
number = {14},
pages = {L205},
author = {Ingo Peschel},
title = {Calculation of reduced density matrices from correlation functions},
journal = {J. Phys. A: Math. Gen.}
}

@article{p-04,
doi = {10.1088/1742-5468/2004/06/P06004},
year = {2004},
month = {6},
publisher = {},
volume = {2004},
number = {06},
pages = {P06004},
author = {Ingo Peschel},
title = {On the reduced density matrix for a chain of free electrons},
journal = {J. Stat. Mech.: Theor. Exp.}
}

@article{pe-09,
doi = {10.1088/1751-8113/42/50/504003},
year = {2009},
month = {12},
publisher = {},
volume = {42},
number = {50},
pages = {504003},
author = {Ingo Peschel and Viktor Eisler},
title = {Reduced density matrices and entanglement entropy in free lattice models},
journal = {J. Phys. A: Math. Theor.}
}

@article{p-12,
	title = {Special review: Entanglement in solvable many-particle models},
	volume = {42},
	issn = {1678-4448},
	doi = {10.1007/s13538-012-0074-1},
	pages = {267--291},
	number = {3},
	journal = {Braz. J. Phys.},
	author = {Peschel, Ingo},
	year = {2012},
}

@article{ac-17,
  title = {Quench action and {R}\'enyi entropies in integrable systems},
  author = {Alba, Vincenzo and Calabrese, Pasquale},
  journal = {Phys. Rev. B},
  volume = {96},
  issue = {11},
  pages = {115421},
  numpages = {8},
  year = {2017},
  month = {9},
  publisher = {American Physical Society},
  doi = {10.1103/PhysRevB.96.115421}
}

@article{fbgvp-21,
	doi = {10.1073/pnas.2106945118},
	year = {2021},
	month = {9},
	publisher = {Proceedings of the National Academy of Sciences},
	volume = {118},
	number = {37},
	author = {Michele Fava and Sounak Biswas and Sarang Gopalakrishnan and Romain Vasseur and S. A. Parameswaran},
	title = {Hydrodynamic nonlinear response of interacting integrable systems},
	journal = {PNAS}
}

@article{agv-19,
  title = {Generalized hydrodynamics, quasiparticle diffusion, and anomalous local relaxation in random integrable spin chains},
  author = {Agrawal, Utkarsh and Gopalakrishnan, Sarang and Vasseur, Romain},
  journal = {Phys. Rev. B},
  volume = {99},
  issue = {17},
  pages = {174203},
  numpages = {12},
  year = {2019},
  month = {5},
  publisher = {American Physical Society},
  doi = {10.1103/PhysRevB.99.174203}
}

@article{bvkm-17,
  title = {Solvable hydrodynamics of quantum integrable systems},
  author = {Bulchandani, Vir B. and Vasseur, Romain and Karrasch, Christoph and Moore, Joel E.},
  journal = {Phys. Rev. Lett.},
  volume = {119},
  issue = {22},
  pages = {220604},
  numpages = {6},
  year = {2017},
  month = {11},
  publisher = {American Physical Society},
  doi = {10.1103/PhysRevLett.119.220604}
}

@article{rc-23,
  doi = {10.48550/ARXIV.2303.01779},
  author = {Rylands, Colin and Calabrese, Pasquale},
  title = {Transport and entanglement across integrable impurities from Generalized Hydrodynamics},
  journal = {arXiv},
  year = {2023}
}

@article{cdlv-18,
  title = {Analytic solution of the domain-wall nonequilibrium stationary state},
  author = {Collura, Mario and De Luca, Andrea and Viti, Jacopo},
  journal = {Phys. Rev. B},
  volume = {97},
  issue = {8},
  pages = {081111},
  numpages = {6},
  year = {2018},
  month = {2},
  publisher = {American Physical Society},
  doi = {10.1103/PhysRevB.97.081111}
}

@article{pdncbf-17,
  title = {Transport in out-of-equilibrium {XXZ} chains: Nonballistic behavior and correlation functions},
  author = {Piroli, Lorenzo and De Nardis, Jacopo and Collura, Mario and Bertini, Bruno and Fagotti, Maurizio},
  journal = {Phys. Rev. B},
  volume = {96},
  issue = {11},
  pages = {115124},
  numpages = {12},
  year = {2017},
  month = {9},
  publisher = {American Physical Society},
  doi = {10.1103/PhysRevB.96.115124},
}

@article{bp-18,
doi = {10.1088/1742-5468/aab04b},
year = {2018},
month = {3},
publisher = {IOP Publishing and SISSA},
volume = {2018},
number = {3},
pages = {033104},
author = {Bruno Bertini and Lorenzo Piroli},
title = {Low-temperature transport in out-of-equilibrium {XXZ} chains},
journal = {J. Stat. Mech.: Theor. Exp.}
}

@article{bpc-18,
  title = {Universal broadening of the light cone in low-temperature transport},
  author = {Bertini, Bruno and Piroli, Lorenzo and Calabrese, Pasquale},
  journal = {Phys. Rev. Lett.},
  volume = {120},
  issue = {17},
  pages = {176801},
  numpages = {6},
  year = {2018},
  month = {4},
  publisher = {American Physical Society},
  doi = {10.1103/PhysRevLett.120.176801},
}

@article{bpp-20,
  title = {Current operators in {B}ethe {A}nsatz and generalized hydrodynamics: An exact quantum-classical correspondence},
  author = {Borsi, M\'arton and Pozsgay, Bal\'azs and Pristy\'ak, Levente},
  journal = {Phys. Rev. X},
  volume = {10},
  issue = {1},
  pages = {011054},
  numpages = {26},
  year = {2020},
  month = {3},
  publisher = {American Physical Society},
  doi = {10.1103/PhysRevX.10.011054}
}

@article{bfpc-18,
doi = {10.1088/1751-8121/aad82e},
year = {2018},
month = {8},
publisher = {IOP Publishing},
volume = {51},
number = {39},
pages = {39LT01},
author = {Bruno Bertini and Maurizio Fagotti and Lorenzo Piroli and Pasquale Calabrese},
title = {Entanglement evolution and generalised hydrodynamics: Noninteracting systems},
journal = {J. Phys. A: Math. Theor.},
}

@article{eb-17,
  title = {Front dynamics and entanglement in the {XXZ} chain with a gradient},
  author = {Eisler, Viktor and Bauernfeind, Daniel},
  journal = {Phys. Rev. B},
  volume = {96},
  issue = {17},
  pages = {174301},
  numpages = {13},
  year = {2017},
  month = {11},
  publisher = {American Physical Society},
  doi = {10.1103/PhysRevB.96.174301}
}

@article{arrs-99,
  title = {Transport in the {XX} chain at zero temperature: Emergence of flat magnetization profiles},
  author = {Antal, T. and R\'acz, Z. and R\'akos, A. and Sch\"utz, G. M.},
  journal = {Phys. Rev. E},
  volume = {59},
  issue = {5},
  pages = {4912--4918},
  numpages = {0},
  year = {1999},
  month = {5},
  publisher = {American Physical Society},
  doi = {10.1103/PhysRevE.59.4912}
}

@article{akr-08,
  title = {Logarithmic current fluctuations in nonequilibrium quantum spin chains},
  author = {Antal, T. and Krapivsky, P. L. and R\'akos, A.},
  journal = {Phys. Rev. E},
  volume = {78},
  issue = {6},
  pages = {061115},
  numpages = {8},
  year = {2008},
  month = {12},
  publisher = {American Physical Society},
  doi = {10.1103/PhysRevE.78.061115}
}

@article{em-18,
  title = {Hydrodynamical phase transition for domain-wall melting in the {XY} chain},
  author = {Eisler, Viktor and Maislinger, Florian},
  journal = {Phys. Rev. B},
  volume = {98},
  issue = {16},
  pages = {161117},
  numpages = {5},
  year = {2018},
  month = {10},
  publisher = {American Physical Society},
  doi = {10.1103/PhysRevB.98.161117}
}

@article{pk-07,
doi = {10.1088/1751-8113/40/8/002},
year = {2007},
month = {2},
publisher = {},
volume = {40},
number = {8},
pages = {1711},
author = {Thierry Platini and Dragi Karevski},
title = {Relaxation in the {XX} quantum chain},
journal = {J. Phys. A: Math. Theor.}
}

@article{adsv-16,
url = {https://dx.doi.org/10.1088/1742-5468/2016/05/053108},
year = {2016},
month = {5},
publisher = {IOP Publishing and SISSA},
volume = {2016},
number = {5},
pages = {053108},
author = {Nicolas Allegra and J\'er\^ome Dubail and Jean-Marie St\'ephan and Jacopo Viti},
title = {Inhomogeneous field theory inside the arctic circle},
journal = {J. Stat. Mech.: Theor. Exp.}
}

@article{sk-23,
  title = {Scaling of fronts and entanglement spreading during a domain wall melting},
  author = {Scopa, Stefano and Karevski, Dragi},
  journal = {arXiv},
  doi = {https://doi.org/10.48550/arXiv.2303.10054},
  year = {2023}
}

@article{vr-16,
doi = {10.1088/1742-5468/2016/06/064007},
year = {2016},
month = {6},
publisher = {IOP Publishing and SISSA},
volume = {2016},
number = {6},
pages = {064007},
author = {Lev Vidmar and Marcos Rigol},
title = {Generalized {G}ibbs ensemble in integrable lattice models},
journal = {J. Stat. Mech.: Theor. Exp.}
}

@article{ef-16,
doi = {10.1088/1742-5468/2016/06/064002},
year = {2016},
month = {6},
publisher = {IOP Publishing and SISSA},
volume = {2016},
number = {6},
pages = {064002},
author = {Fabian H L Essler and Maurizio Fagotti},
title = {Quench dynamics and relaxation in isolated integrable quantum spin chains},
journal = {J. Stat. Mech.: Theor. Exp.}
}

@article{eip-09,
doi = {10.1088/1742-5468/2009/02/P02011},
year = {2009},
month = {2},
publisher = {},
volume = {2009},
number = {02},
pages = {P02011},
author = {Viktor Eisler and Ferenc Iglói and Ingo Peschel},
title = {Entanglement in spin chains with gradients},
journal = {J. Stat. Mech.: Theor. Exp.}
}

@article{gkk-02,
title = {{W}annier–{S}tark resonances in optical and semiconductor superlattices},
journal = {Phys. Rep.},
volume = {366},
number = {3},
pages = {103-182},
year = {2002},
issn = {0370-1573},
doi = {https://doi.org/10.1016/S0370-1573(02)00142-4},
author = {Markus Glück and Andrey R. Kolovsky and Hans Jürgen Korsch}
}

@book{gp-13,
  title={Solid State Physics},
  author={Grosso, G. and Parravicini, G.P.},
  year={2000},
  publisher = {Academic Press},
  address = {London},
  doi = {10.1016/B978-0-12-304460-0.X5000-2},
}

@article{w-62,
  title = {Dynamics of Band Electrons in Electric and Magnetic Fields},
  author = {Wannier, Gregory H.},
  journal = {Rev. Mod. Phys.},
  volume = {34},
  issue = {4},
  pages = {645--655},
  numpages = {0},
  year = {1962},
  month = {10},
  publisher = {American Physical Society},
  doi = {10.1103/RevModPhys.34.645}
}

@article{hkkm-04,
	title = {Dynamics of {B}loch oscillations},
	volume = {6},
	issn = {1367-2630},
	doi = {10.1088/1367-2630/6/1/002},
	pages = {2--2},
	journal = {New J. Phys.},
	author = {Hartmann, T and Keck, F and Korsch, H J and Mossmann, S},
	date = {2004-01-14},
}

@article{b-29,
	title = {\"Uber die {Q}uantenmechanik der {E}lektronen in {K}ristallgittern},
	volume = {52},
	issn = {0044-3328},
	doi = {10.1007/BF01339455},
	pages = {555--600},
	number = {7},
	journal = {Zeitschrift für Physik},
	author = {Bloch, Felix},
	date = {1929-07-01},
}

@article{mlbcfkpygm-21,
	title = {Observation of {S}tark many-body localization without disorder},
	volume = {599},
	issn = {1476-4687},
	doi = {10.1038/s41586-021-03988-0},
	pages = {393--398},
	number = {7885},
	journal = {Nature},
	author = {Morong, W. and Liu, F. and Becker, P. and Collins, K. S. and Feng, L. and Kyprianidis, A. and Pagano, G. and You, T. and Gorshkov, A. V. and Monroe, C.},
	date = {2021-11-01},
}

@article{shmp-19,
  title = {Stark many-body localization},
  author = {Schulz, M. and Hooley, C. A. and Moessner, R. and Pollmann, F.},
  journal = {Phys. Rev. Lett.},
  volume = {122},
  issue = {4},
  pages = {040606},
  numpages = {5},
  year = {2019},
  month = {1},
  publisher = {American Physical Society},
  doi = {10.1103/PhysRevLett.122.040606},
}

@article{nbr-19,
    author = {Evert van Nieuwenburg  and Yuval Baum  and Gil Refael },
    title = {From {B}loch oscillations to many-body localization in clean interacting systems},
    journal = {PNAS},
    volume = {116},
    number = {19},
    pages = {9269-9274},
    year = {2019},
    doi = {10.1073/pnas.1819316116},
}

@article{gghzy-21,
	title = {Observation of {B}loch oscillations and {W}annier-{S}tark localization on a superconducting quantum processor},
	volume = {7},
	issn = {2056-6387},
	doi = {10.1038/s41534-021-00385-3},
	pages = {51},
	number = {1},
	journal = {npj Quantum Information},
	author = {Guo, Xue-Yi and Ge, Zi-Yong and Li, Hekang and Wang, Zhan and Zhang, Yu-Ran and Song, Pengtao and Xiang, Zhongcheng and Song, Xiaohui and Jin, Yirong and Lu, Li and Xu, Kai and Zheng, Dongning and Fan, Heng},
	date = {2021-03-19},
}

@article{dgp-21,
	title = {Stark many-body localization: Evidence for {H}ilbert-space shattering},
	volume = {103},
	issn = {2469-9950, 2469-9969},
	doi = {10.1103/PhysRevB.103.L100202},
	pages = {L100202},
	number = {10},
	journal = {Phys. Rev. B},
	author = {Doggen, Elmer V. H. and Gornyi, Igor V. and Polyakov, Dmitry G.},
	date = {2021-03-04}
}

@article{lzsg-19,
  title = {Quasilocalized excitations induced by long-range interactions in translationally invariant quantum spin chains},
  author = {Lerose, Alessio and \ifmmode \check{Z}\else \v{Z}\fi{}unkovic, Bojan and Silva, Alessandro and Gambassi, Andrea},
  journal = {Phys. Rev. B},
  volume = {99},
  issue = {12},
  pages = {121112},
  numpages = {6},
  year = {2019},
  month = {3},
  publisher = {American Physical Society},
  doi = {10.1103/PhysRevB.99.121112}
}

@article{yy-69,
  doi = {10.1063/1.1664947},
  title={Thermodynamics of a one-dimensional system of bosons with repulsive delta-function interaction},
  author={Yang, Chen-Ning and Yang, Cheng P},
  journal={J. Math. Phys.},
  volume={10},
  number={7},
  pages={1115--1122},
  year={1969},
  publisher={American Institute of Physics}
}

@book{n-98-book,
  doi = {https://doi.org/10.1201/9780429497926},
  author = {Negele, J.W.},
  title = {Quantum Many-particle Systems},
  publisher = {CRC Press},
  year = {1998}
}

@misc{NIST:DLMF,
    title = {{NIST Digital Library of Mathematical Functions}},
    note = {Release 1.1.6 of 2022-06-30},
    year = {2022},
    url = {http://dlmf.nist.gov/},
    editor = {F.~W.~J. Olver and A.~B. {Olde Daalhuis} and D.~W. Lozier and B.~I. Schneider and R.~F. Boisvert and C.~W. Clark and B.~R. Miller and B.~V. Saunders and H.~S. Cohl and M.~A. McClain}
}

@article{f-17,
  title = {Higher-order generalized hydrodynamics in one dimension: The noninteracting test},
  author = {Fagotti, Maurizio},
  journal = {Phys. Rev. B},
  volume = {96},
  issue = {22},
  pages = {220302},
  numpages = {6},
  year = {2017},
  month = {12},
  publisher = {American Physical Society},
  doi = {10.1103/PhysRevB.96.220302}
}

@article{ce-10,
doi = {10.1088/1742-5468/2010/08/P08029},
year = {2010},
month = {8},
publisher = {},
volume = {2010},
number = {08},
pages = {P08029},
author = {Pasquale Calabrese and Fabian H L Essler},
title = {Universal corrections to scaling for block entanglement in spin-1/2 {XX} 
chains},
journal = {J. Stat. Mech.: Theor. Exp.},
}

@article{kl-09,
  title = {Quantum noise as an entanglement meter},
  author = {Klich, Israel and Levitov, Leonid},
  journal = {Phys. Rev. Lett.},
  volume = {102},
  issue = {10},
  pages = {100502},
  numpages = {4},
  year = {2009},
  month = {3},
  publisher = {American Physical Society},
  doi = {10.1103/PhysRevLett.102.100502}
}

@article{cmv-12,
url = {https://dx.doi.org/10.1209/0295-5075/98/20003},
year = {2012},
month = {4},
publisher = {},
volume = {98},
number = {2},
pages = {20003},
author = {Pasquale Calabrese and Mihail Mintchev and Ettore Vicari},
title = {Exact relations between particle fluctuations and entanglement in Fermi gases},
journal = {Europhys. Lett.},
}

@article{cmc-23,
  title={Full counting statistics and symmetry resolved entanglement for free conformal theories with interface defects}, 
  author={Luca Capizzi and Sara Murciano and Pasquale Calabrese},
  year={2023},
  eprint={2302.08209},
  doi={https://doi.org/10.48550/arXiv.2302.08209}  
}

@article{mplcg-19,
  title = {Suppression of transport in nondisordered quantum spin chains due to confined excitations},
  author = {Mazza, Paolo Pietro and Perfetto, Gabriele and Lerose, Alessio and Collura, Mario and Gambassi, Andrea},
  journal = {Phys. Rev. B},
  volume = {99},
  issue = {18},
  pages = {180302},
  numpages = {6},
  year = {2019},
  month = {5},
  publisher = {American Physical Society},
  doi = {10.1103/PhysRevB.99.180302}
}

@article{lsmpcg-20,
  title = {Quasilocalized dynamics from confinement of quantum excitations},
  author = {Lerose, Alessio and Surace, Federica M. and Mazza, Paolo P. and Perfetto, Gabriele and Collura, Mario and Gambassi, Andrea},
  journal = {Phys. Rev. B},
  volume = {102},
  issue = {4},
  pages = {041118},
  numpages = {7},
  year = {2020},
  month = {7},
  publisher = {American Physical Society},
  doi = {10.1103/PhysRevB.102.041118}
}

\end{document}